\documentclass{aa}
\usepackage{graphicx}
\usepackage{txfonts}
\usepackage[dvipsnames]{xcolor}
\usepackage[breaklinks=true]{hyperref}
 \hypersetup{
     colorlinks=true,
     linkcolor=blue,
     filecolor=blue,
     citecolor=blue,      
     urlcolor=blue}

\begin{document}

   %\title{Think about a nice title...}
   \title{Progenitors of low-mass binary black-hole mergers in the isolated binary evolution scenario}
   %\subtitle{I. subtitle maybe not needed...}

   \author{Federico Garc\'ia \inst{1,2}
          \and
          Adolfo Simaz Bunzel \inst{3} \thanks{Fellow of CONICET}
          \and
          Sylvain Chaty \inst{1,4}
          \and
          Edward Porter \inst{4}
          \and
          Eric Chassande-Mottin \inst{4}
          }

   \institute{AIM, CEA, CNRS, Universit\'e Paris-Saclay, Universit\'e de Paris, F-91191 Gif-sur-Yvette France
     \and
     Kapteyn Astronomical Institute, University of Groningen, PO Box 800, NL-9700 AV Groningen, the Netherlands
     \and
     Instituto Argentino de Radioastronom\'ia, Universidad Nacional de La Plata, 1900 La Plata, Argentina
     \and
     Universit\'e de Paris, CNRS, Astroparticule et Cosmologie, F-75013 Paris, France
   }

   \date{received ... 2020 ; accepted ... 2021}

% \abstract{}{}{}{}{} 
% 5 {} token are mandatory
 
  \abstract
   {The formation history, progenitor properties and expected rates of the binary black holes discovered by the LIGO-Virgo collaboration, through the gravitational-wave emission during their coalescence, are now a topic of active research.}
   {We aim to study the progenitor properties and expected rates of the two lowest-mass binary black hole mergers, GW\,151226 and GW\,170608, detected within the first two Advanced LIGO-Virgo observing runs, in the context of the classical isolated binary-evolution scenario.} 
   {We use the publicly-available 1D-hydrodynamic stellar-evolution code {\tt MESA}, which we adapted to include the black-hole formation and the unstable mass transfer developed during the so-called common-envelope phase. Using more than $60\,000$ binary simulations, we explore a wide parameter space for initial stellar masses, separations, metallicities and mass-transfer efficiencies. We obtain the expected distributions for the chirp mass, mass ratio and merger time delay by accounting for the initial stellar binary distributions. We predict the expected merger rates that we compare with the detected gravitational-wave events, and study the dependence of our predictions with respect to (yet) unconstrained parameters inherent to binary stellar evolution.}
   {Our simulations for both events show that, while the progenitors we obtain are compatible over the entire range of explored metallicities, they show a strong dependence on the initial masses of the stars, according to stellar winds. All the progenitors found follow a similar evolutionary path, starting from binaries with initial separations in the $30-200$ R$_\odot$ range, experiencing a stable mass transfer interaction before the formation of the first black hole, and a second unstable mass-transfer episode leading to a common-envelope ejection that occurs either when the secondary star crosses the Hertzsprung gap or when it is burning He in its core. The common-envelope phase plays a fundamental role in the considered low-mass range: only progenitors experiencing such an unstable mass-transfer phase are able to merge in less than a Hubble time.} 
   {We find integrated merger-rate densities in the range 0.2--5.0~yr$^{-1}$~Gpc$^{-3}$ in the local Universe for the highest mass-transfer efficiencies explored. The highest rate densities lead to detection rates of 1.2-3.3 yr$^{-1}$, being compatible with the observed rates. The common-envelope efficiency $\alpha_{\rm CE}$ has a strong impact on the progenitor populations. A high-efficiency scenario with $\alpha_{\rm CE}=2.0$ is favoured when comparing the expected rates with observations.}


   \maketitle
%
%________________________________________________________________

\section{Introduction}
\label{sec:intro}
In 2015, the Advanced LIGO \citep{aligo} and Advanced Virgo \citep{avirgo} collaboration (LVC) began a series of observation runs.  During both the O1 (September 12, 2015 - January 19, 2016) and O2 (November 30, 2016 - August 25, 2017) observation runs, a total of 11 gravitational wave (GW) events were observed.  Ten of these events were the detection of signals from the merger of binary black holes \citep[BBHs,][]{2019PhRvX...9c1040A} and one corresponded to the merger of two neutron stars \citep[GW170817]{2017PhRvL.119p1101A}.

While these BBH are mainly dominated by high-mass components ($M \ga 35\,M_{\odot}$), two detections in particular, GW151226 \citep{2016PhRvL.116x1103A} and GW170608 \citep{2017ApJ...851L..35A}, are low-mass systems having BH masses consistent with those found in X-ray binaries (i.e. $M \la 20\,M_{\odot}$).  Despite that all these events could belong to the same population \citep{2019ApJ...882L..24A}, the existence and abundance of these objects trigger the question of their formation history. Several scenarios have been proposed in the literature, including the isolated binary evolution  \citep[our main focus here,][]{1998ApJ...506..780B, 2016MNRAS.458.2634M, 2017ApJ...846..170T} and the dynamical formation channels \citep{2000ApJ...528L..17P, 2014MNRAS.440.2714B, 2016PhRvD..93h4029R}.

In the dynamical formation scenario, BBHs are produced by three-body encounters in stellar clusters. In the chemically homogeneous evolutionary channel, compact BBHs are formed from rapidly rotating stars in near contact binaries that experience efficient internal mixing.
% which only works for massive stars in low metallicity environments where strong losses of angular-momentum due to winds are not present.
It is estimated that dynamical encounters in globular clusters contributed to less than a few percent of all observed events \citep{2014MNRAS.440.2714B, 2016PhRvD..93h4029R}, while BBH rates in open and young star clusters can be an order of magnitude higher \citep{2019MNRAS.487.2947D, 2020arXiv200110690K}. On the other hand, the chemically homogeneous scenario is not able to produce BBH in low-mass range, with $M \lesssim 10$~M$_{\odot}$ \citep{2016A&A...588A..50M}. In this study, we concentrate on the classical isolated binary evolution scenario where the formation of the ultra-compact binary leading to the BBH merger is driven through an unstable mass-transfer phase where a common-envelope (CE) is ejected \citep{2013A&ARv..21...59I,2016A&A...596A..58K}.

Our main goal is to study the progenitor population of the lightest BBHs detected by Advanced LIGO and Advanced Virgo during their first two science runs, O1 and O2, and its dependence on the uncertainties intrinsically related to binary stellar evolution such as the accretion efficiency during a stable mass-transfer phase, efficiency of the CE ejection, impact of metallicity and the evolution of merger rates with redshift.

This kind of studies has been usually performed following a binary population synthesis approach using several different numerical codes \citep{1997AstL...23..492L,2002ApJ...572..407B,2003MNRAS.342.1169V,2016Natur.534..512B,2016MNRAS.462.3302E,2017NatCo...814906S,2018MNRAS.481.1908K,2019MNRAS.485..889S}. In this work, we perform detailed numerical stellar simulations of the binary systems, using the 1D-hydrodynamic stellar-evolution code {\tt MESA}. Such kind of treatment, which has been recently growing \citep[see, for instance,][]{2017A&A...604A..55M}, allows for an accurate modelling of the mass-transfer between the binary components, that has consequences on the final BH masses before the merger. However, the method is computationally expensive, which is the reason why it is usually not considered in standard binary population studies. Our simulations incorporate the evolution during the CE phase, which plays a fundamental role in the considered low-mass BBH range.

The paper is organised as follows: we first describe the binary stellar evolution using MESA in Section \ref{sec:MESA}, we then focus on the results for GW\,170608 and GW\,151226 in Section \ref{sec:results}, before reporting on the population-weighted results in Section \ref{sec:PW-results}, and giving the projected merger and gravitational-wave event rates in Section \ref{sec:merger-GW-event-rates}. We finally discuss the results in Section \ref{sec:discussion}, and summarise and conclude this paper in Section \ref{sec:summary-conclusions}.

\section{Binary stellar evolution using {\tt MESA}}
\label{sec:MESA}

Here, we present models of stellar-binary systems that evolve starting from zero-age main sequence (ZAMS), to the formation of binary black holes (BBH) and their final merger through the emission of gravitational waves (GW). We made use of the publicly-available stellar evolution code, {\tt MESA} \citep{2019ApJS..243...10P,2018ApJS..234...34P,2013ApJS..208....4P,2011ApJS..192....3P,2010ascl.soft10083P}, which we modified\footnote{\url{https://github.com/asimazbunzel/mesa_low_mass_bbhs}} to include a treatment for the common-envelope (CE) phase, BH formation, and to properly merge in a single run the three evolutionary stages involved in this problem, that is: a binary of massive stars, massive stellar evolution and BH formation, and the formation of a binary BH.

\subsection{Microphysics, nuclear networks and stellar winds}
\label{sec:stellar-evolution}

Our simulations are computed using {\tt MESA} version 10398. We use CO-enhanced opacity tables from the OPAL project \citep{Iglesias1993, Iglesias1996}. Convection is modelled following the standard mixing-length theory \citep[MLT, ][]{1958ZA.....46..108B} adopting a mixing-length parameter $\alpha_{\rm{MLT}}=1.5$. Convective regions are determined using the Ledoux criterion. In late evolutionary stages of massive stars, the convective velocities in certain regions of the convective envelope can approach the speed of sound, running out of the domain of applicability of the MLT. For these regions, we use an MLT++ treatment \citep{2013ApJS..208....4P} that reduces the super-adiabaticity. Semi-convection is included according to the diffusive approach presented in \cite{1983A&A...126..207L} which depends on an efficiency parameter that we assume to be $\alpha_{\rm{SC}}=1.0$. We also include a convective-core overshooting during H burning extending the core radius given by the Ledoux criterion by 0.335 of the pressure-scale height \citep[$H_P$, ][]{2011A&A...530A.116B}. When mass is transferred from one star to its companion, the material accreted by the accretor may have a mean molecular weight higher than its outer layers. This leads to an unstable situation that induces a thermohaline mixing \citep{1980A&A....91..175K}, which is included by adopting $\alpha_{\rm{th}}=1.0$. In this work, we only consider non-rotating stars, and hence we ignore the effects that tidal interactions may have on internal rotation and mixing, and their impact on final BH masses \citep{2000ApJ...528..368H}.

We use standard thermonuclear reaction networks provided by {\tt MESA}: {\tt basic.net} for the hydrogen and helium burning phases, and switch during run time to {\tt co\_burn.net} for the carbon burning phase. Furthermore, stellar winds are modelled using mass-loss rates depending on effective temperatures and surface H mass fraction ($X_{\rm{s}}$). When $T_{\rm{eff}}>10^{4}\;\rm{K}$, for $X_{\rm{s}}>=0.4$ we use the prescription from \cite{2001A&A...369..574V}, while for $X_{\rm{s}}<0.4$ we apply that from \cite{2000A&A...360..227N}. When $T_{\rm{eff}}<10^{4}\;\rm{K}$, we adopt the prescription from \cite{1988A&AS...72..259D}.

\subsection{From stellar binaries to binary black holes}
\label{sec:binary-evolution}

Stellar binaries and their interactions are modelled using the {\tt MESAbinary} module of {\tt MESA}. Our simulations start when both stars with masses $M_{{\rm i},1}$ and $M_{{\rm i},2}$ are at the zero-age main sequence (ZAMS), in circular orbits, at a certain initial separation $a_{\rm i}$.

The mass-exchange between the two binary components is modelled as follows. To determine which star is the donor or the accretor, the atmospheric transfer (MT) rates of both stars is compared according to \cite{1988A&A...202...93R}. When one of the stars overfills its Roche lobe (RLO), we apply an implicit MT scheme to obtain the MT rate ($\dot{M}_{\rm{RLOF}}$) at each step. The MT stability is controlled as described in Sec.~\ref{sec:CE}.

The accretion efficiency, $\epsilon$, is assumed to remain constant throughout the entire evolution, and only considers the mass lost through an isotropic wind in the vicinity of the accretor. Assuming no mass loss from either direct fast winds or a circumbinary co-planar toroid, the efficiency of MT ($\epsilon=1-\beta$) is defined through the $\beta$ parameter from {\tt MESAbinary}, which is equal to the fraction of transferred mass that is isotropically lost with the angular momentum of the accretor. Hence, $\epsilon=0$ means fully-inefficient mass transfer (i.e. no accretion).

Once the first BH is formed in the system, we use the {\tt point mass} approximation from {\tt MESAbinary} and we limit the accretion onto the compact object to a factor of the Eddington rate 
 %\begin{equation}
    $\dot{M}_{\rm Edd, BH} = 4 \pi G M_{\rm BH} / \eta \kappa_{\rm donor}$, 
%\end{equation}
where $G$ is the gravitational constant, $\kappa_{\rm donor}$ is the opacity of the donor star at its surface and $\eta$ is the radiation efficiency of the BH which we set to 1\% implying super-Eddington accretion. The change in the orbital angular momentum is inferred from the effects of mass loss in the binary (MT and stellar winds).

If a second BH forms, thus leading to a BBH, the time to merger, $t_{\rm merger}$, is estimated from \cite{1964PhRv..136.1224P} based on the component masses ($M_{\rm BH}$), their mutual separation and eccentricity.

\subsection{Black hole formation}
\label{sec:bh-formation}

When a non-degenerate star completes the carbon core burning phase, its evolution is stopped as the binary parameters will not change appreciably during the later evolutionary stages due to their short duration \citep{2006csxs.book..623T}. The BH formation is modelled according to \cite{2012ApJ...749...91F}. Given the actual CO core mass, and the expected BH remnant mass obtained by the {\em delayed} collapse prescription, we update the orbital parameters of the binary immediately after BH formation, following \cite{1991PhR...203....1B}:

\begin{equation} \label{app:bh-formation-dynamics}
    a_{\rm post-SN} = \dfrac{\mu}{2\mu - 1} a_{\rm pre-SN} , \\
    e_{\rm post-SN} = \dfrac{1 - \mu}{\mu}
\end{equation}
where 
\begin{equation}
    \mu = \dfrac{M_{{\rm post-SN}, 1}+M_{{\rm post-SN}, 2}}{M_{{\rm pre-SN}, 1}+M_{{\rm pre-SN}, 2}}.
\end{equation}

At this point, we also check for disruption at BH formation, given by $e_{\rm post-SN} > 1$.    
%assuming an instantaneous, spherically symmetric ejection of the H and He stellar envelopes. 
In this calculation, we neglect any interaction with the binary companion and do not consider asymmetric kicks (for a discussion on the impact of asymmetric kicks onto our results, see Appendix~\ref{app:kicks}).
%To update the orbital parameters of the binary immediately after BH formation, we follow \cite{1991PhR...203....1B} and assume that the mass of the BH remnant is 80\% of the stellar CO core mass. The remaining 20\% accounts for the release of gravitational binding energy.
%We also check for possible binary disruption, considering the following parameters: the mass ejected, the final separation and eccentricity. %While this recipe for BH formation and compact object final mass is clearly not unique, it appears efficient for the modelling of binary evolution. 
%Core-collapse and compact object formation is a highly-debated topic \citep[see, for instance,][and references therein]{2012ApJ...749...91F,2016ApJ...821...38S}. 
%A comparison between our prescription and others present in the literature can be found in Appendix~\ref{app:bh_formation}.

\subsection{Common-envelope phase}
\label{sec:CE}

\subsubsection{Definition}

A common-envelope (CE) phase occurs when the MT becomes unstable. The stability of MT in binary systems is usually understood in terms of the reaction of the binary components to mass accretion or loss \citep{1997A&A...327..620S}. Binary population synthesis (BPS) codes generally associate the MT stability to the binary mass ratio at the onset of the MT phase. If this ratio is above some limit, then the MT is considered unstable thus, typically, leading to a CE phase.

However, this was recently revised in \cite{2015MNRAS.449.4415P} and \cite{2017MNRAS.465.2092P}, showing that the mass ratio condition is not sufficient, nor necessary, to predict the outcome of the MT phase. In these papers based on numerical stellar evolution, the authors show that binaries with mass ratio\footnote{Here $m_1$ and $m_2$ are the masses of the donor and accretor at the onset of RLO, respectively.} $q=m_2/m_1$ as low as $0.13$ experience a stable MT phase, contradicting earlier works \citep[e.g.,][]{2008ApJS..174..223B}.

%Based on these recent works we introduce new conditions for deciding the MT stability using the rate of mass transfer $\dot{M}_{\rm RLOF}$ during the RLO at different evolutionary stages. 
In this work
%When the binary consists of two non-degenerate stars, 
we assume the MT to be unstable when, during RLO, the MT rate exceeds a certain value that we fix to the Eddington limit of the donor, 
%\begin{equation} \label{eq:mdot_edd}
    $\dot{M}_{\rm Edd} = 4 \pi c R / \kappa$,
%\end{equation}
where $R$ is the stellar radius and $c$ is the speed of light. When the binary consists of two non-degenerate stars, we also consider unstable MT if the MT rate is higher than the Eddington limit of the accretor. In our simulations these MT rates are of the order of $\sim 10^{-2}$~M$_\odot$~yr$^{-1}$, consistent with the value assumed by \citet{2019A&A...628A..19Q} for unstable MT. In contrast to population synthesis codes, {\tt MESA} allows us to calculate the MT rate at each evolutionary time step. It is thus possible to continuously verify these conditions, even when the binary experiences a RLO phase. This additionally allows us to detect late phases of unstable MT rates in the case of long and initially stable RLO phases.

When any of the above conditions are met, a so-called CE phase is triggered. During this phase, the donor star engulfs its companion, while the accretor in-spirals inside the envelope of the donor. A successful envelope ejection may occur on a dynamical timescale \citep{2001ASPC..229..239P}. The CE phase plays a crucial role in reducing the separation between two stars, or between a star and a BH, in a binary system, by a factor of 10 to 100 \citep{2017ApJ...846..170T}, thus producing {\em ultra-compact} BBHs. This is fundamental, since no BBH is expected to merge in less than the Hubble time when the post-CE system is not ultra-compact in nature, in the case when no asymmetric kicks are considered.

\subsubsection{Numerical implementation}

In order to implement a numerical treatment for the CE phase within {\tt MESA}, we use the so-called energy formalism \citep{1984ApJ...277..355W, 1990ApJ...358..189D}. According to this formalism, the main energy source needed to eject the stellar envelope is provided by the orbital energy reservoir and thus, by the in-spiral of the companion. Changes in these two quantities are related by a free parameter $\alpha_{\rm CE}$ representing the fraction of the orbital energy deposited as kinetic energy of the envelope components:
\begin{equation}\label{eq:CE}
    \Delta E_{\rm{bind}} = \alpha_{\rm{CE}} \: \Delta E_{\rm{orb}} , 
\end{equation}
where $\Delta E_{\rm bind}$ is the change in the binding energy of the donor star envelope, while $\Delta E_{\rm{orb}}$ represents the released orbital energy throughout the in-spiral, and $\alpha_{\rm CE}$ is the CE efficiency that we assume to be fixed throughout the entire CE phase. Here, $E_{\rm bind}$ is given by
\begin{equation}
    E_{\rm bind} = \int_{M_{\rm core}}^{M} \left( -\dfrac{G m_r}{r} + u \right) {\rm d}m_{\rm r}
\end{equation}
which includes both the gravitational binding energy and the specific internal energy of the envelope. The latter has an additional term associated to the recombination energy of available H and He, known to help with the ejection of the envelope \citep{2015MNRAS.447.2181I, 2015MNRAS.450L..39N, 2016A&A...596A..58K}. 

Given an unstable MT rate $\dot{M}$, during the time step $\Delta t$, the donor losses a mass $\Delta M = \dot{M} \Delta t$ from its outer layer, changing its envelope binding energy by $\Delta E_{\rm bind}$, and consequently the orbital energy by $\Delta E_{\rm orb}$, which naturally leads to the spiral-in of the binary.
%=- G M M'/2a$ (where $G$ is the gravitational constant, $M$ is the mass of the donor, $M'$ the mass of the accretor and $a$ their separation. Hence, the actual changes in $\Delta E_{\rm orb}$ and $\Delta M$ lead to a new (lower) orbital separation: that is, the binary spiral-in.

For numerical stability reasons, once a CE phase is triggered, during a fixed amount of time (that we set to 10~yr), we linearly increase the MT rate up to a fixed maximum value (that we set to $10^{-1}$~M$_\odot$~yr$^{-1}$ throughout this work), we assume that BH mass grow is negligible during this relatively short episode, and thus we turn off mass accretion onto the companion \citep{2015ApJ...803...41M, 2020ApJ...897..130D}.
In \citet{2020A&A...636A.104B}, the authors argue that recent calculations show that the accretion rates onto compact objects in CE inspiral can be reduced even by a factor of $\sim$100 with respect to Bondi-Hoyle accretion when the structure of the envelope is taken into account. Moreover, for most density gradients considered by \citet{2017ApJ...838...56M}, the accretion rate is well below 10\% of Bondi-Hoyle accretion rate. Based on these findings, \citet{2020A&A...636A.104B} conclude that BHs of $\sim$30~M$_\odot$ accrete $\sim$0.5~M$_\odot$ in a typical CE event. In our case, considering BHs of $\sim$10~M$_\odot$ would lead to an even lower mass accretion during a CE, which is well within the uncertainties of the BH masses in the considered GW events (see Section~\ref{sec:results}).

Once the maximum value for the MT rate is reached, we keep that value constant until the donor star detaches, i.e., its radius becomes smaller than its corresponding Roche lobe, or until the merger of the two stars becomes unavoidable, i.e., the envelope could not be successfully ejected, leading to a single star or a so-called Thorne-Zytkow object \citep[TZO, ][]{1977ApJ...212..832T}. In this latter case, the evolution is stopped as it would not lead to a BBH\footnote{We assume that a merger happens in a binary when the reduction in separation leads to a relative donor overflow $r_{\rm{RL}} = \dfrac{R-R_{\rm{RL}}}{R_{\rm{RL}}}$ bigger than a limiting value which we set equal to 20. We found that beyond this value the donor radius cannot become smaller than its corresponding Roche lobe. Additionally, we assume a merger occurs when the simulations would not complete due to convergence issues at late times during the CE phase.}. In the former case, when reaching the detach condition, the mass transfer rate $\dot{M}_{\rm RLOF}$ is linearly decreased, as a fraction of the radius of the lobe, down to the mass loss $\dot{M}_{\rm th}$ obtained at thermal equilibrium. For those surviving binaries, the donor star becomes an almost naked core, with a tiny envelope rich in H, and with a close companion. The evolution then returns to the standard {\tt MESA} workflow, allowing for a new stable RLO phase to start.

\subsection{{\tt MESA} runs}
\label{sec:mesa-runs}

As our main goal is to study the progenitor population of the lightest BBHs detected by the LVC during the O1/O2 runs, we explore a wide range of metallicities, i.e. $Z= 0.0001$, 0.001, 0.004, 0.007 and 0.015, which, in principle, can lead to BHs in the mass range of interest. In addition, in order to study the dependence of our results on the poorly-known MT efficiency, we cover a wide range for this parameter with four different values: $\epsilon= 0.6$, 0.4, 0.2 and 0.0, going from efficient to fully inefficient regimes. For each pair of Z and $\epsilon$ values, we compute a 3D grid in the parameter space formed by the initial masses ($M_{\rm{i},1}, M_{\rm{i},2}$) and the binary initial separation ($a_{\rm{i}}$).

As a first approach, we fix\footnote{We refer the reader to \cite{2013A&ARv..21...59I} for a complete discussion on values of CE efficiency parameter $\alpha_{\rm{CE}} \geq 1.0$.} $\alpha_{\rm{CE}} = 2.0$, and explore a wide range of initial separations that lead to interacting binaries, from $30-4000$~R$_\odot$ with relatively large logarithmic step of 0.03~dex. In this initial inspection, we found binaries that went through a CE phase when the first BH was formed for systems with $a_{\rm{i}} < 500$~R$_\odot$. Once the broad parameter space was understood, we focused our searches on this evolutionary channel, but exploring wider ranges of masses, according to metallicity, and lowering the grid spacing in $a_{\rm i}$ to 0.02 dex, for values below $500$~R$_\odot$, in order to constrain the regions containing actual solutions, which we call {\em target regions}.

As each {\tt MESA} simulation is computationally expensive, from this point onward, we set up a strategy to concentrate our runs on the regions that lead to BBHs with masses in the range of interest, avoiding the calculation of binary systems leading to too light or too heavy chirp masses, but also systems that did not display strong interaction (high MT rates) and thus led to extremely long merging times. These ranges depend mainly on $Z$, but also on $\epsilon$. Thus, for each parameter combination, our runs were set up to cover different ranges, in an iterative fashion, until the target regions were finally bounded. 

In order to explore the dependence of our results on the CE efficiency, we ran another set of simulations with $\alpha_{\rm{CE}} = 1.0$. In this case, we only ran those simulations for which we already had found a CE trigger and chirp masses in the range of interest. Since the density of CE survivals significantly decreases for this efficiency, due to a natural increase in CE mergers, we decided to increase the grid resolution to $\Delta M = 1$~M$_\odot$ and 0.01 dex in the logarithmic grid of $a_{\rm i}$, and we proceeded to run the 26 first neighbours in the refined grid for each CE survival of the initial runs. After this step we proceeded in an iterative manner surrounding the next family of survivals and so-on until the process converged. 

In Table~\ref{table_runs} we summarise the $66\,632$ simulations computed using our {\tt MESA}-based numerical code for each MT efficiency $\epsilon$ and CE efficiency $\alpha_{\rm CE}$ explored. Full details of the parameter space explored for this work are presented in Appendix~\ref{app:mesa_runs} and a full example of a typical {\tt MESA} simulation leading to a BBH formation after a CE phase is shown in Appendix~\ref{app:example}.

\begin{table}[]
\caption{Number of {\tt MESA} runs performed for this work}
\centering
\begin{tabular}{lrr}
\hline
$\epsilon$ & $\alpha_{\rm CE} = 2.0$ & $\alpha_{\rm CE} = 1.0$ \\
\hline

0.6     & 10644                &  9568                \\
0.4     & 10530                & 12269                \\
0.2     & 10490                &  8665                \\
0.0     &  4466                &   --                 \\

\hline
Total   & 36130                & 30502                 \\            
\hline
\end{tabular}
\label{table_runs}
\end{table}

\section{Results for GW\,151226 and GW\,170608}
\label{sec:results}

The response of detectors such as Advanced LIGO and Advanced Virgo to a GW compact binary coalescence depends not only on the distance and relative orientation of the GW source to the detector, but also on the intrinsic binary properties; the most important being the chirp mass ($\mathcal{M}_{\rm chirp}$), which affects the phase evolution of gravitational waveform \citep{1996PhRvD..53.2878F} and is defined as,
    $\mathcal{M}_{\rm chirp} = \mu^{3/5} M^{2/5}$
where $\mu = M_{\rm{BH},1}\,M_{\rm{BH},2}/(M_{\rm{BH},1}+M_{\rm{BH},2})$ is the reduced mass and $M=M_{\rm{BH},1}+M_{\rm{BH},2}$ the total mass of the BBH.

GW detections can be used to infer measurements of the redshifted chirp mass in the detector frame, i.e. $\mathcal{M}_{\rm chirp}^{\rm detector} = (1+z)\mathcal{M}_{\rm chirp}^{\rm source}$. In order to estimate $\mathcal{M}_{\rm chirp}$ in the source frame, and hence to be able to compare with our theoretical results, the binary masses have to be un-redshifted.  Unfortunately, while a direct measurement of the luminosity distance can be made from an inspiral event, without an electromagnetic counterpart, a cosmological model has to be assumed to extract the redshift of the source. For this work, we used a flat $\Lambda$CDM model with $H_0=70$~km~s$^{-1}$ and $T_{\rm CMB}=2.725$~K and the {\tt astropy.cosmology} package \citep{2013A&A...558A..33A,2018AJ....156..123A} to estimate the masses of the lowest-mass BBHs detected by Advanced~LIGO-Virgo in O1/O2 runs \citep{2019PhRvX...9c1040A}. We found $\mathcal{M}_{\rm chirp} = 8.83^{+0.74}_{-0.66}$~M$_{\odot}$ and $q_{\rm BBH} = 0.56^{+0.44}_{-0.49}$ for GW151226 and $\mathcal{M}_{\rm chirp} = 7.91^{+0.43}_{-0.37}$~M$_{\odot}$ and $q_{\rm BBH} = 0.69^{+0.31}_{-0.56}$ for GW170608, respectively, in their 100\% confidence intervals (C.I.).

%We found $\mathcal{M}_{\rm chirp} = 8.83^{+0.40}_{-0.33}$ and $q_{\rm BBH} = 0.56^{+0.41}_{-0.37}$ for GW151226 and $\mathcal{M}_{\rm chirp} = 7.91^{+0.22}_{-0.22}$ and $q_{\rm BBH} = 0.69^{+0.29}_{-0.41}$ for GW170608, respectively, in their 95\% confidence intervals (C.I.).

Throughout this work we consider a certain binary model to be a possible progenitor compatible with any of the GW events under study if its $\mathcal{M}_{\rm chirp}$ and $q_{\rm BBH}$ lays within the 100\% C.I. of the corresponding GW event and if it has a merger time delay ($t_{\rm merger}$) shorter than the Hubble time ($\tau_{\rm Hubble} = 13.46$~Gyr, under our cosmological assumptions).

\subsection{Parameter space and target regions}
\label{sec:parameter-space}

We use models and method described in Sec.~\ref{sec:MESA}, in order to find the target region in the 3D parameter space, associated with each GW event, for each metallicity, MT and CE efficiencies.

\begin{figure}
   \centering
        \includegraphics[width=\hsize]{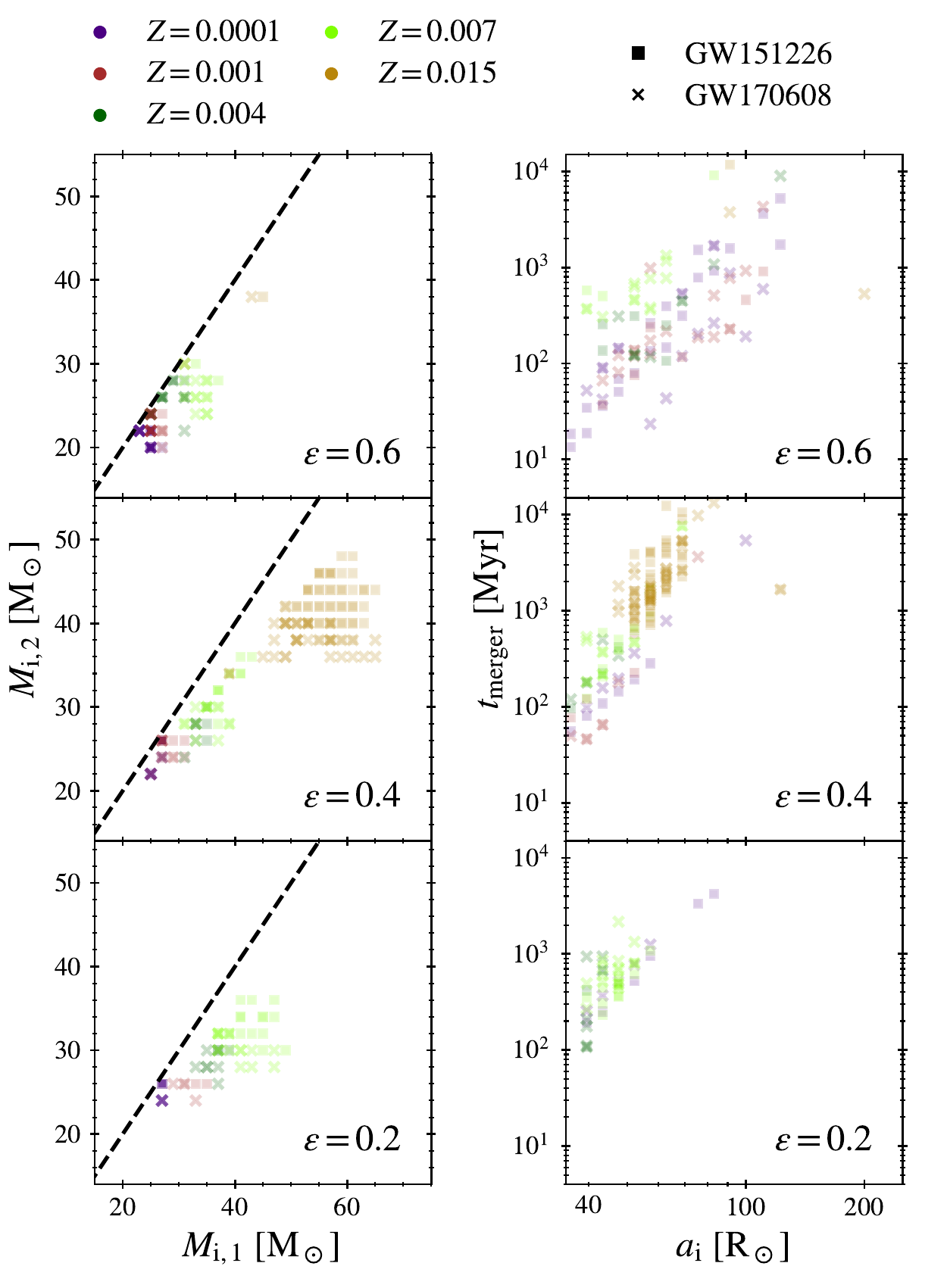}
            \caption{Target regions of the parameter space for GW151226 (square markers) and GW170608 (cross markers) for models with $\alpha_{\rm{CE}}=2.0$. On the left panels we show the progenitor initial masses ($M_{i,1}>M_{i,2}$), while on the right panels we plot the merger time delay ($t_{\rm merge}$) against initial binary separation ($a_i$). Panels from top to bottom correspond to each set of efficiencies: $\epsilon=0.6$, $0.4$, $0.2$. Dashed lines indicate equal progenitor masses.}
         \label{fig:parameter-space-a2}
\end{figure}

\begin{figure}
   \centering
        \includegraphics[width=\hsize]{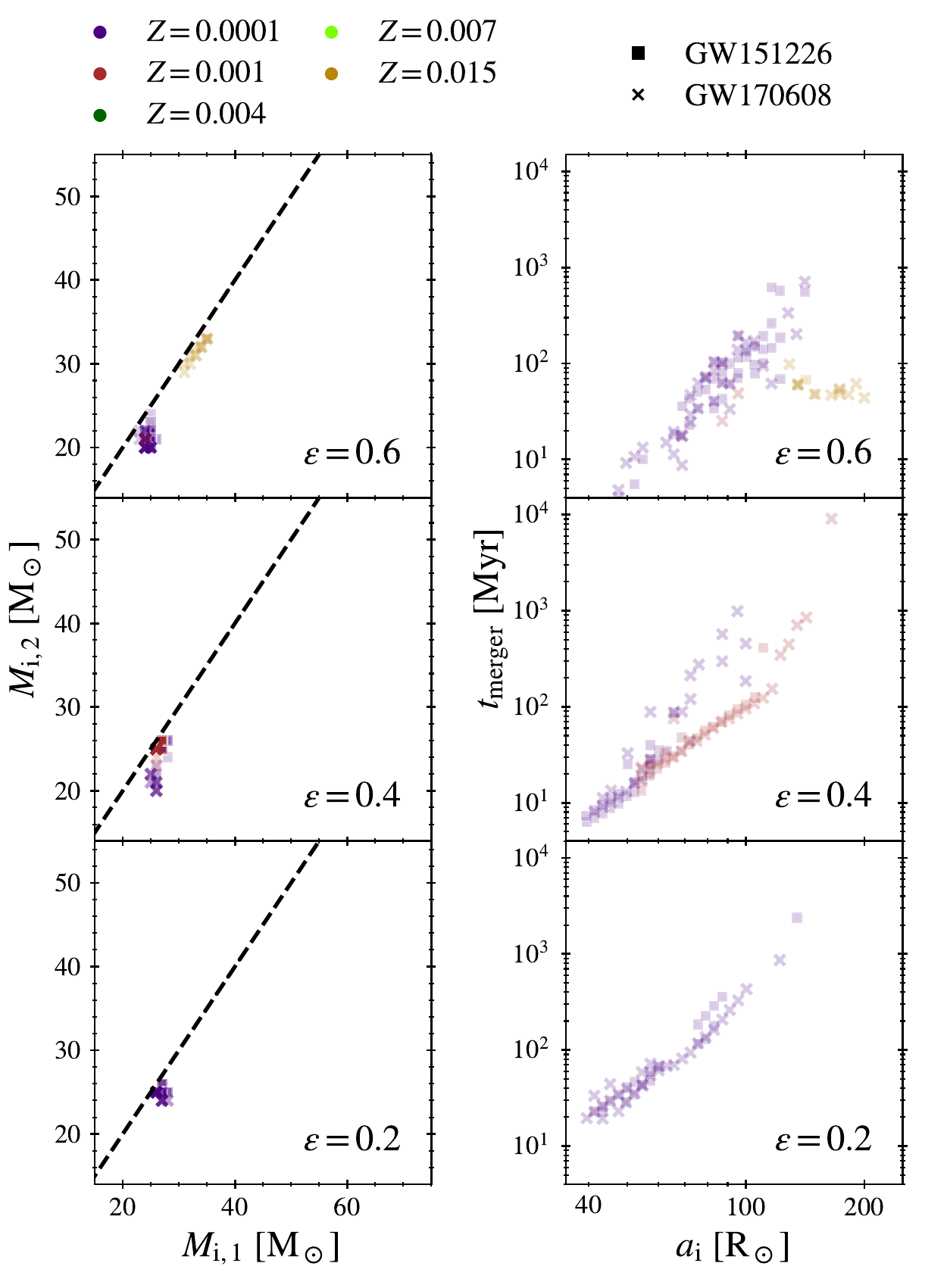}
            \caption{Idem to Figure \ref{fig:parameter-space-a2} for $\alpha_{\rm CE}=1.0$. More points are obtained as a result of increasing the grid resolution in the parameter space.}  %Note that we omitted the lowest efficiencies $\epsilon \leq 0.2$ since no BBH solutions compatible with GW151226 nor GW170608 were found.}
         \label{fig:parameter-space-a1}
\end{figure}

Figure~\ref{fig:parameter-space-a2} shows the solution regions for GW151226 and GW170608 obtained with $\alpha_{\rm CE}=2.0$, along with their merger time delay. In general, more massive progenitors are needed to explain GW151226 than GW170608 in each individual case, in agreement with their final BH masses. Higher metallicities require increasingly massive stars in order to obtain progenitors of both GW events, a direct consequence of the dependence of stellar winds on the metallicity content \citep[see for instance][]{kudritzki:2000}. We find that this effect is independent of the MT efficiency. For all metallicities explored at the higher MT efficiencies $\epsilon \geq 0.4$, we find progenitors compatible with both GW events, while in the $\epsilon=0.2$ MT regime, we find that only binaries with $Z \le 0.007$ are able to become actual progenitors. This is because in the $\epsilon=0.2$ and $Z=0.015$ case, the BBH that merge within a Hubble time have chirp masses below the lower boundaries given by \cite{2019PhRvX...9c1040A} for the least massive BBHs. Furthermore, no compatible progenitors are found for the fully-inefficient MT scenario (i.e. $\epsilon=0$).

Another interesting feature in Figure~\ref{fig:parameter-space-a2} is that, at low metallicities, when $Z \leq 0.004$ and high MT efficiencies $\epsilon \geq 0.4$, binaries with similar initial masses are admissible progenitors.
% These binaries result from the high accretion efficiencies during the interaction between the two non-degenerate stars,  two lowest MT efficiencies.
Efficient accretion favours the growth of a convective core in the accreting star, which in our case is typically located on the main-sequence (MS), leading to a rejuvenation \citep{1995A&A...297..483B, 2007MNRAS.376...61D}, and thus a longer duration of the core H-burning phase that can delay the H depletion after the primary (and initially more massive) star collapses to a BH.

However, this behaviour is not observed at high metallicities as rejuvenation is not strong enough to delay H depletion. In this case, after an initial efficient MT phase, the secondary star expands after leaving the MS and both stars overfill their Roche lobes, evolving to an over-contact phase. This, in principle, is not the same as a CE phase as co-rotation can be maintained as long as there is no overflow through the second Lagrangian point (L2) and thus no viscous drag as in the CE phase. Although our simulation does not allow for an over-contact phase, it is expected that BHs produced by this channel have higher masses than the ones found for GW151226 and GW170608 \citep{2016A&A...588A..50M}. Combining this last line of reasoning with strong winds, we find no solutions for high metallicities and low MT efficiencies.

In the right hand side panels of Figure~\ref{fig:parameter-space-a2}, it can be seen that, for all MT efficiencies, increasing $a_{\rm i}$ leads to increased merger time delay. In addition, values of $t_{\rm merger}$ cover up to two orders of magnitude for a given $a_{\rm i}$. This is explained by a larger scatter in BH masses at the BBH formation stage, since separations and eccentricities remain less spread after the second BH has formed.

Figure~\ref{fig:parameter-space-a1} shows the target regions found for $\alpha_{\rm CE}=1.0$. We see that the progenitors have mass ratios close to unity. The rest of the progenitors obtained with a lower mass ratio either merge during the CE phase or produce BBHs outside the boundaries in $\mathcal{M}_{\rm chirp}$ and $q_{\rm BBH}$. Low-metallicity progenitors are preferred in all cases, but a family of high-metallicity progenitors is found in the highest MT efficiency scenario (i.e. $\epsilon = 0.6$), with relatively high initial separations ($a_i \sim 100-200$~R$_\odot$). The latter are not present for $\alpha_{\rm CE}=2$, as they do not merge within a Hubble time.

The solutions obtained for $\alpha_{\rm CE}=2.0$ with high metallicities and short initial separations $a_{\rm i}<80$~R$_\odot$ merge during the CE phase and thus do not produce BBHs. Additionally, those binaries which end up being compatible progenitors for $\alpha_{\rm CE}=1.0$ have smaller separations at BBH formation than their respective $\alpha_{\rm CE}=2.0$ runs, and thus they all have their associated $t_{\rm merger}$ effectively reduced.

\subsection{Black hole masses}

\begin{figure}
    \centering
    \includegraphics[width=\hsize]{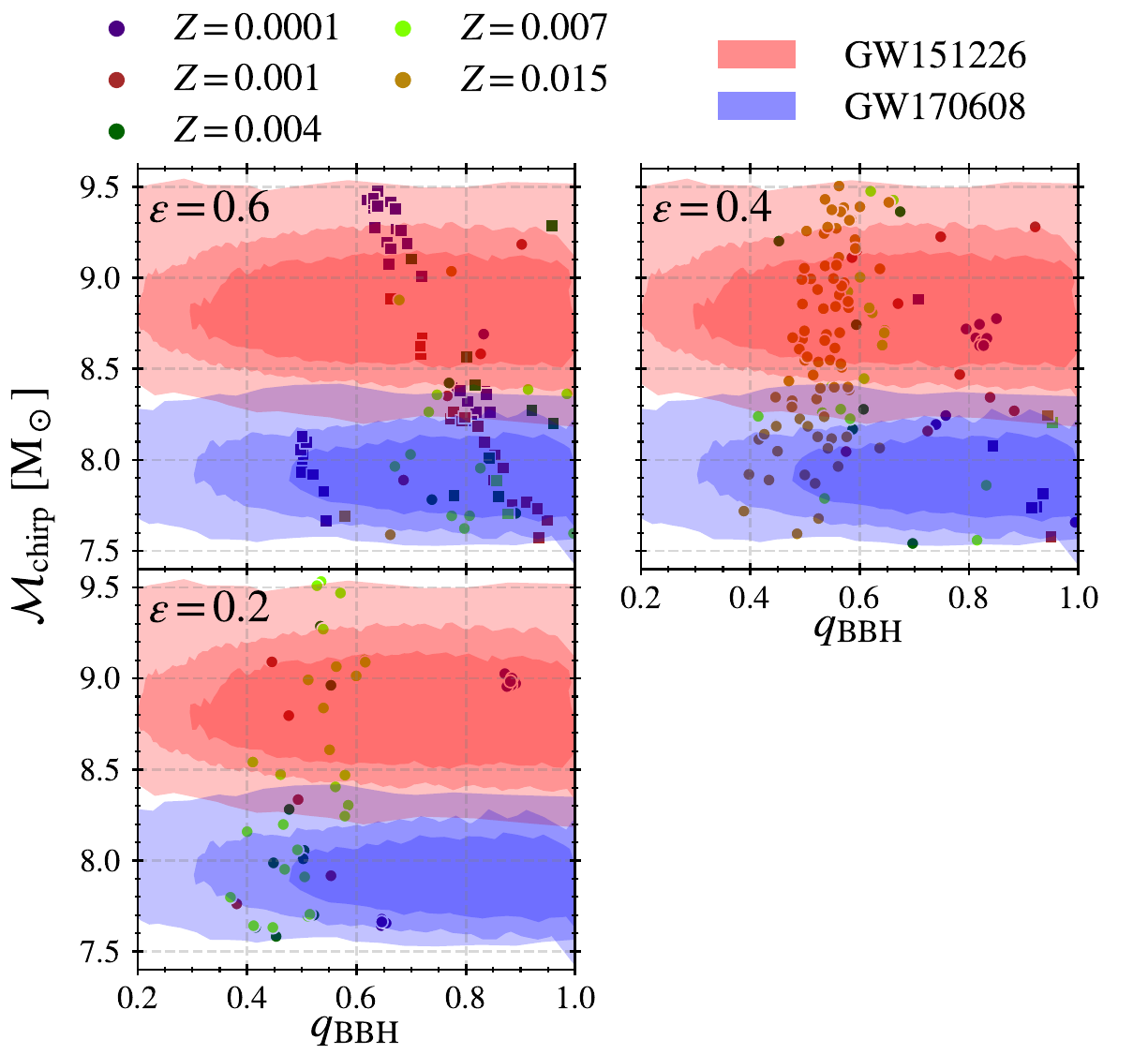}
    \caption{Chirp masses ($\mathcal{M}_{\rm chirp}$) and mass ratios ($q_{\rm BBH}$) of BBHs compatible with GW151226 and/or GW170608 events (within their 100\%, 90\% and 68\% credible intervals in salmon and blue shaded areas, respectively), that merge within the Hubble time for $\alpha_{\rm CE}=2.0$. Each panel corresponds to a different value of the MT efficiency. Square (round) markers correspond to binaries with $M_{\rm{BH},2}>M_{\rm{BH},1}$ ($M_{\rm{BH},2}<M_{\rm{BH},1}$). Different point colours correspond to each metallicity adopted in this work (see legend).
    }
    \label{fig:q-mchirp-a2}
\end{figure}

\begin{figure}
    \centering
    \includegraphics[width=\hsize]{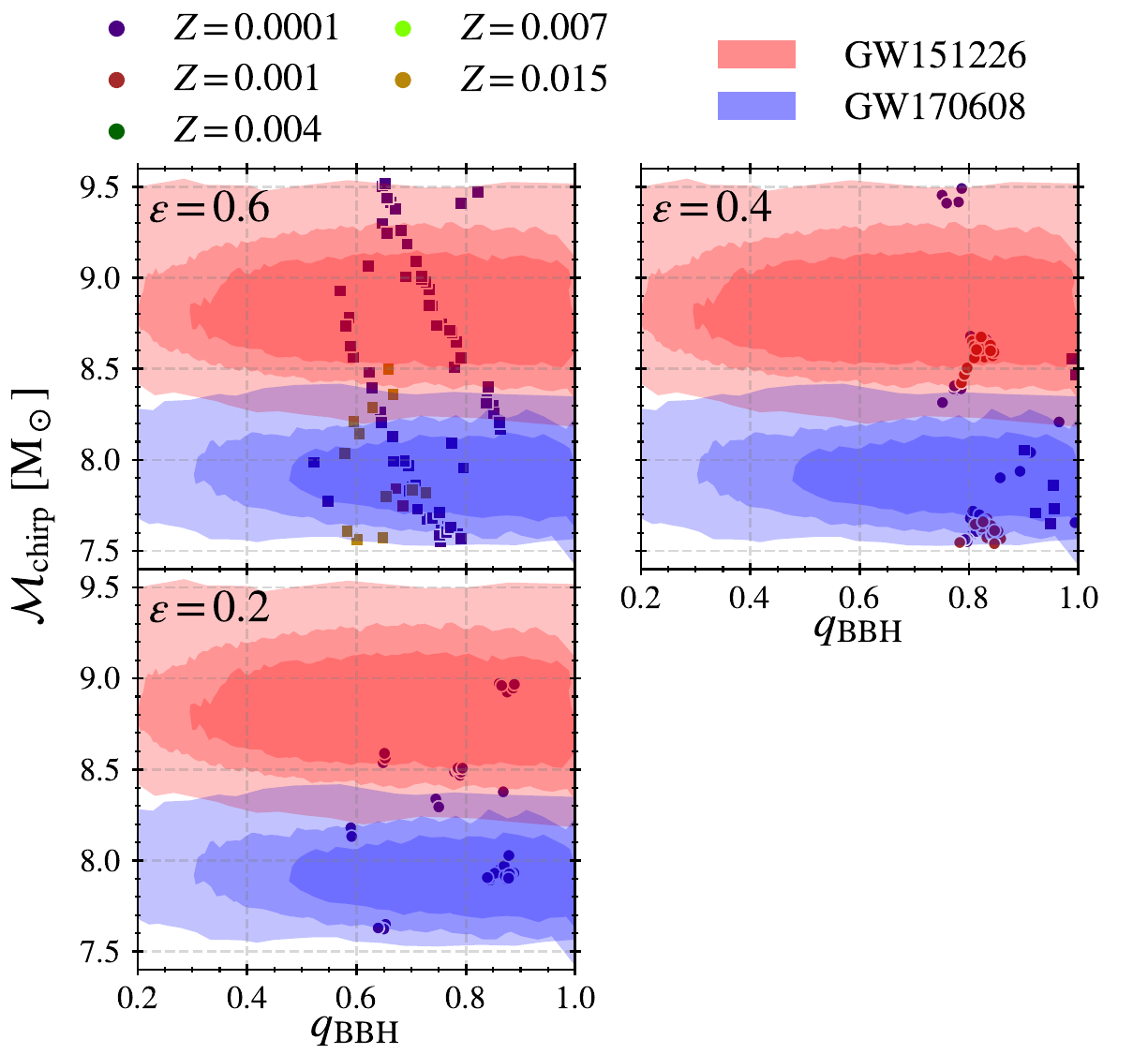}
    \caption{Idem to Figure~\ref{fig:q-mchirp-a2} for $\alpha_{\rm CE}=1.0$.} %No BBH solutions compatible with GW151226 nor with GW170608 are found at low efficiencies $\epsilon \leq 0.2$, and are ommitted here.}
    \label{fig:q-mchirp-a1}
\end{figure}

In Figure~\ref{fig:q-mchirp-a2} we present the distribution of BH masses associated with the progenitors found using $\alpha_{\rm CE}=2.0$. Independently of the MT efficiency, the binaries have $q_{\rm BBH} \ga 0.4$ and cover the entire range in $\mathcal{M}_{\rm chirp}$. Large MT efficiencies (such as $\epsilon = 0.6$) tend to form BBHs with mass ratios closer to unity, while low MT efficiencies (such as $\epsilon = 0.2$) tend to form BBHs with unequal-mass BHs ($q_{\rm BBH} \approx 0.4-0.6$). The intermediate scenario (such as $\epsilon = 0.4$) can form BBHs with a broad range of mass ratios, depending on the metallicity.

BBHs obtained at lower metallicities span the entire range of mass ratios, while $0.4 \lesssim q_{\rm BBH}  \lesssim 0.7$ for the higher metallicities as a consequence of the high mass-loss rates associated with stellar winds. Interestingly, in the latter range of metallicities, for some cases (showed in square markers in Figure~\ref{fig:q-mchirp-a2}), the most massive BH is formed last due to the rejuvenation of the secondary (and initially least massive) star during the stable MT stage. Additionally, at low metallicities, such binaries concentrate along lines of decreasing $\mathcal{M}_{\rm chirp}$ when $q_{\rm BBH}$ increases.

In Figure~\ref{fig:q-mchirp-a1} we present BH mass properties obtained with $\alpha_{\rm CE}=1.0$. All BBHs have $q_{\rm BBH} \ga 0.5$. When $\epsilon=0.6$, BBHs can also be formed at the highest metallicity. Almost all of them went through a rejuvenation process which produced a secondary BH more massive than the primary. On the other hand, while for $\epsilon=0.4$ this is only achieved for the lowest metallicity, in the case $\epsilon=0.2$, this is never the case.

\subsection{Merger time delay}

\begin{figure}
    \centering
    \includegraphics[width=\hsize]{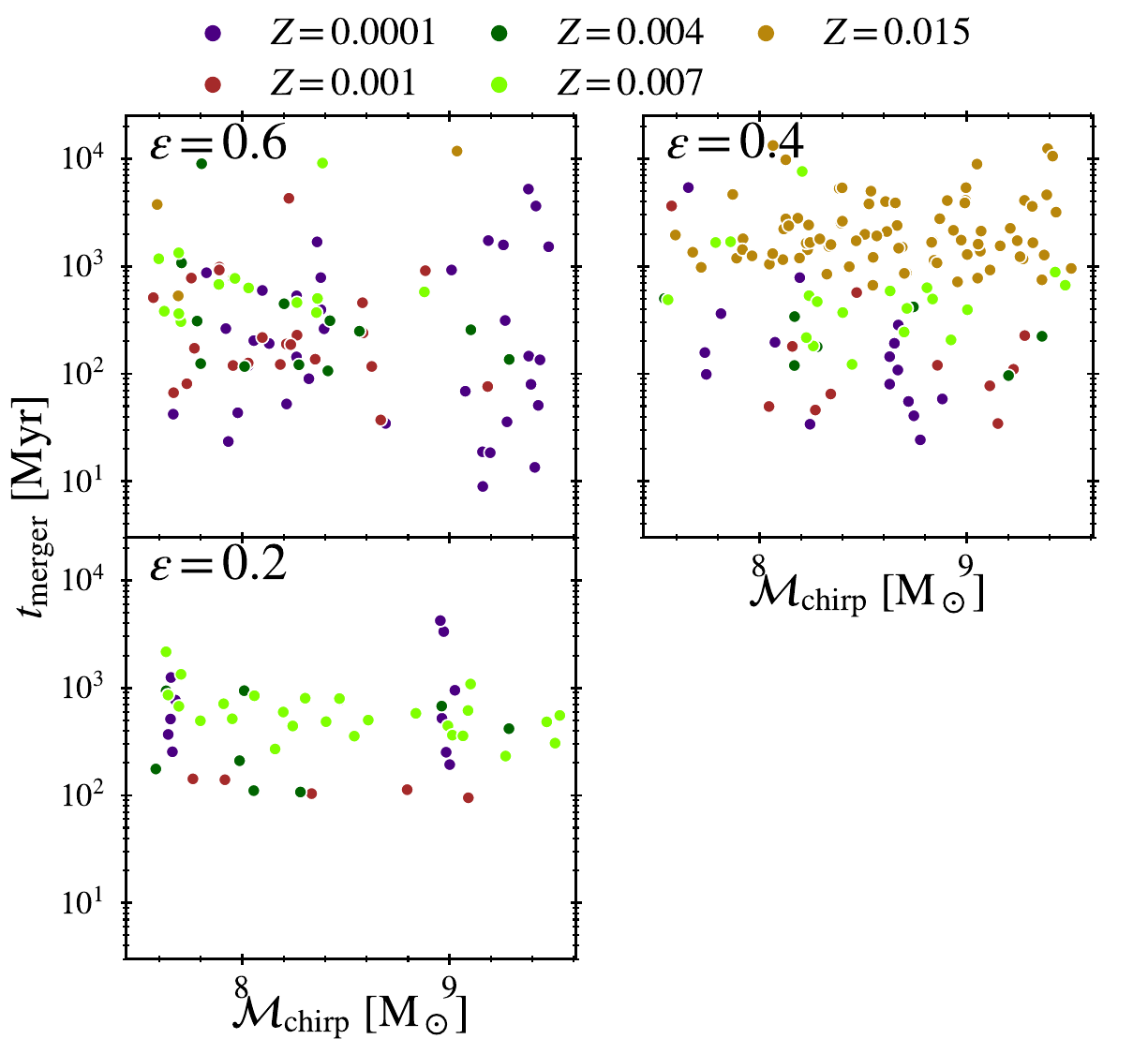}
    \caption{Merger time delay ($t_{\rm merger}$) of BBHs compatible with GW151226 and/or GW170608 (within their 100\% C.I.) for $\alpha_{\rm CE}=2.0$. Each panel corresponds to different values of the MT efficiency. Different point colours correspond to each metallicity adopted in this work (see legend).}
    \label{fig:merger-times-a2}
\end{figure}

\begin{figure}
    \centering
    \includegraphics[width=\hsize]{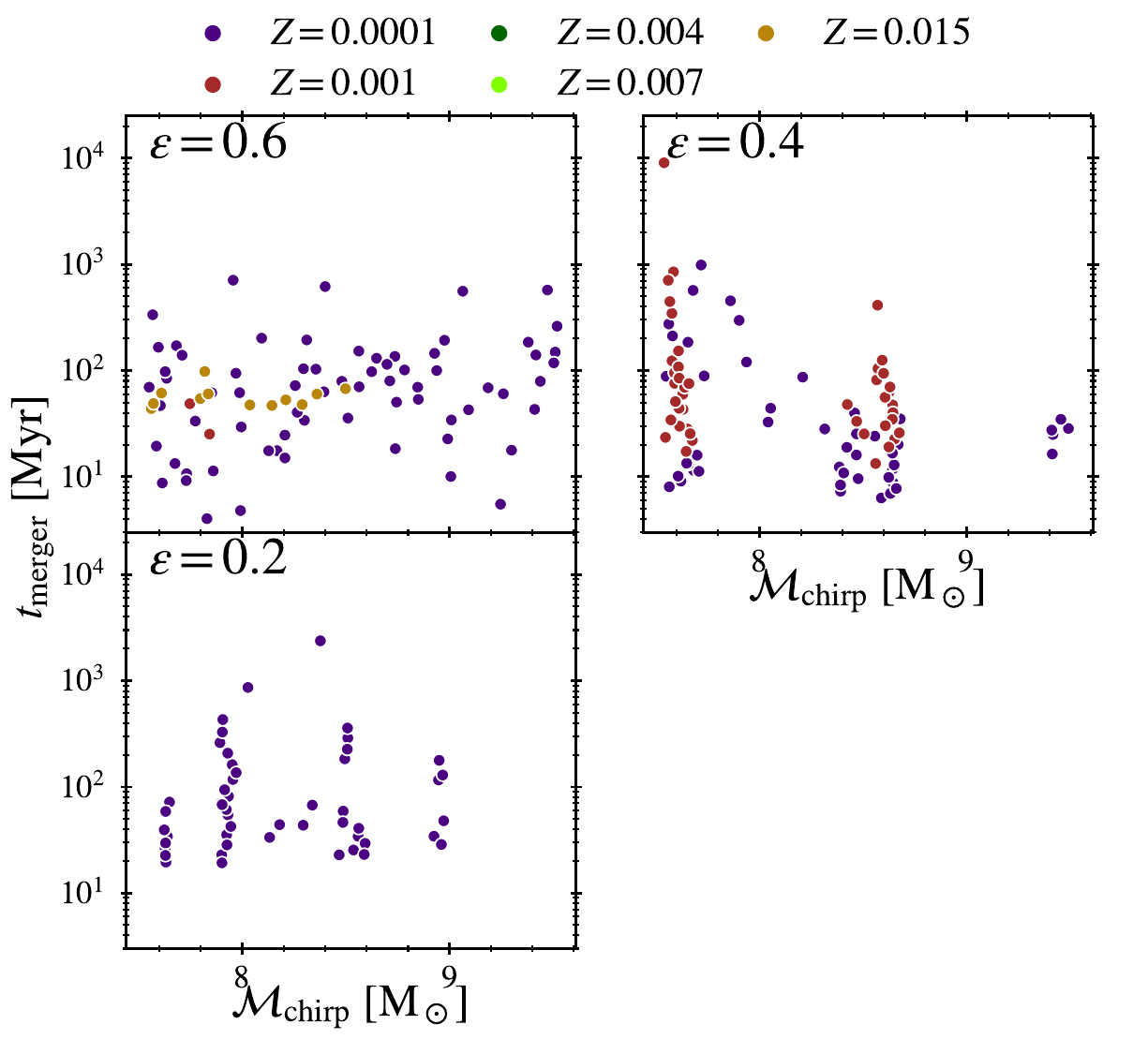}
    \caption{Same as Figure~\ref{fig:merger-times-a2} for $\alpha_{\rm CE}=1.0$.} % No BBH solutions compatible with GW151226 nor GW170608 are found for the lowest efficiencies $\epsilon \leq 0.2$ which are thus omitted here.}
    \label{fig:merger_times-a1}
\end{figure}

In Figures~\ref{fig:merger-times-a2} and \ref{fig:merger_times-a1} we present the distribution of merger time delay $t_{\rm merger}$ as a function of $\mathcal{M}_{\rm chirp}$ for all BBHs with masses compatible to GW151226 and/or GW170608, for $\alpha_{\rm CE}=2.0$ and $\alpha_{\rm CE}=1.0$ respectively. When  $\alpha_{\rm CE}=2.0$ BBHs merge after long delays $t_{\rm merger} \sim 0.1-10$~Gyr, comparable to Hubble time, while, when  $\alpha_{\rm CE}=1.0$ the mergers occur with shorter delays, $t_{\rm merger} \la 1$~Gyr, and typically 10--100~Myr.

Delay times play a fundamental role in determining the age of the stellar population from which the observed BBHs originate. The results above imply that in the former set of simulations, old binary systems are more involved, while in the latter, younger binary-system progenitors are favoured. However, in this case, high metallicities are strongly disfavoured (except for the highest MT efficiency, $\epsilon=0.6$), setting strong constraints on the expected properties of their possible host galaxies. Although the contribution of asymmetric natal kicks could change these distributions.

Interestingly, for all simulated binaries, the CE phase is required for the binary to merge within a Hubble time, and merger time delays are strongly impacted by the assumed CE efficiency. As expected, the CE phase plays a fundamental role in BBH mergers in the isolated binary channel. More details can be found in Appendix~\ref{app:time_delay}.

\section{Metallicity-dependent weighted population}

\label{sec:PW-results}

The results obtained so far rely over regularly grids that uniformly sample the space of initial masses and separations. In this section, we produce metallicity-dependent population-weighted results, re-scaling by empirical initial mass functions (IMF) for the primary and secondary stars and by an initial separation distribution computed from the observed binary orbital period $\mathcal{P}$ distributions.

\subsection{Assumptions and methodology}

For the mass  $M_{i,1}$ of the primary and initially most massive star, we use the IMF from \cite{1993MNRAS.262..545K}
\begin{equation}
    \xi(M) \propto
            \begin{cases} 
                (M/M_0)^{-\alpha_0} & M_{\rm{low}} \leq M < M_0 \\
                (M/M_0)^{-\alpha_1} & M_0 \leq M < M_1 \\
                (M_1/M_0)^{-\alpha_1}\,(M/M_1)^{-\alpha_2} & M_1 \leq M \leq M_\mathrm{high} 
            \end{cases}
\end{equation}
where $\alpha_0=1.3$, $\alpha_1=2.2$ and $\alpha_3=2.7$, while $M_\mathrm{low}=0.08$~M$_\odot$, $M_0=0.5$~M$_\odot$, $M_1=1$~M$_\odot$ and $M_\mathrm{high}=150$~M$_\odot$.

Given $M_{i,1}$, the mass  $M_{i,2}$ of the secondary star is drawn from a flat distribution in the mass ratio $q = M_{i,2}/M_{i,1}$,
\begin{equation}
    \xi(q) = \dfrac{1}{q_{\rm{max}} - q_{\rm{min}}},
\end{equation}
where $q_{\rm{min}}=0.1$ and $q_{\rm{max}}=1.0$.

The initial separation is drawn from the orbital period distribution given in \cite{2012Sci...337..444S} and \cite{2015ApJ...814...58D},
\begin{equation}
    \xi(\mathcal{P}) \propto \mathcal{P}^{-0.55}, \quad \text{for~} 0.15 < \mathcal{P} < 5.5
\end{equation}
where $\mathcal{P} = \log P_{\rm{orb}}$ in units of days. We note that when drawing the separations from this distribution we have assumed zero eccentricity to keep consistency with our simulations.

Although orbital properties seem to be relatively unaffected by metallicity in the range between the Milky Way and the Large Magellanic Cloud metallicities \citep{2017A&A...598A..84A}, throughout this work we assume that these distributions are preserved for the entire range of metallicities.

For each metallicity $Z$ and MT efficiency $\epsilon$, we randomly draw $10^{7}$ binaries from the distributions described above. To get a reasonable resolution, we restrict the random draws to the relevant ranges of masses and separations and keep track of the normalisation constant to account for the rest of the distributions, that is otherwise ignored in the Monte-Carlo simulation.

A massive star binary corresponds to a point in the parameter space defined by $M_{\rm{i},1}, M_{\rm{i},2}$ and $a_{\rm i}$. This point is mapped to the closest point in the regular grid introduced in Sec.~\ref{sec:mesa-runs}. We assign to the randomly generated binary the properties of the closest binary evolved through the {\tt MESA} simulations presented in Sec.~\ref{sec:MESA}.

This method allows us to obtain statistics representative of the entire binary star population, based on numerical simulations of binary stellar evolution. Such a treatment is usually not considered in standard population synthesis simulations as it is computationally expensive.

\subsection{Population-weighted results for GW151226 and GW170608}

\subsubsection{Properties of the initial binaries}

\begin{figure*}
    \centering
    \includegraphics[width=0.9\textwidth]{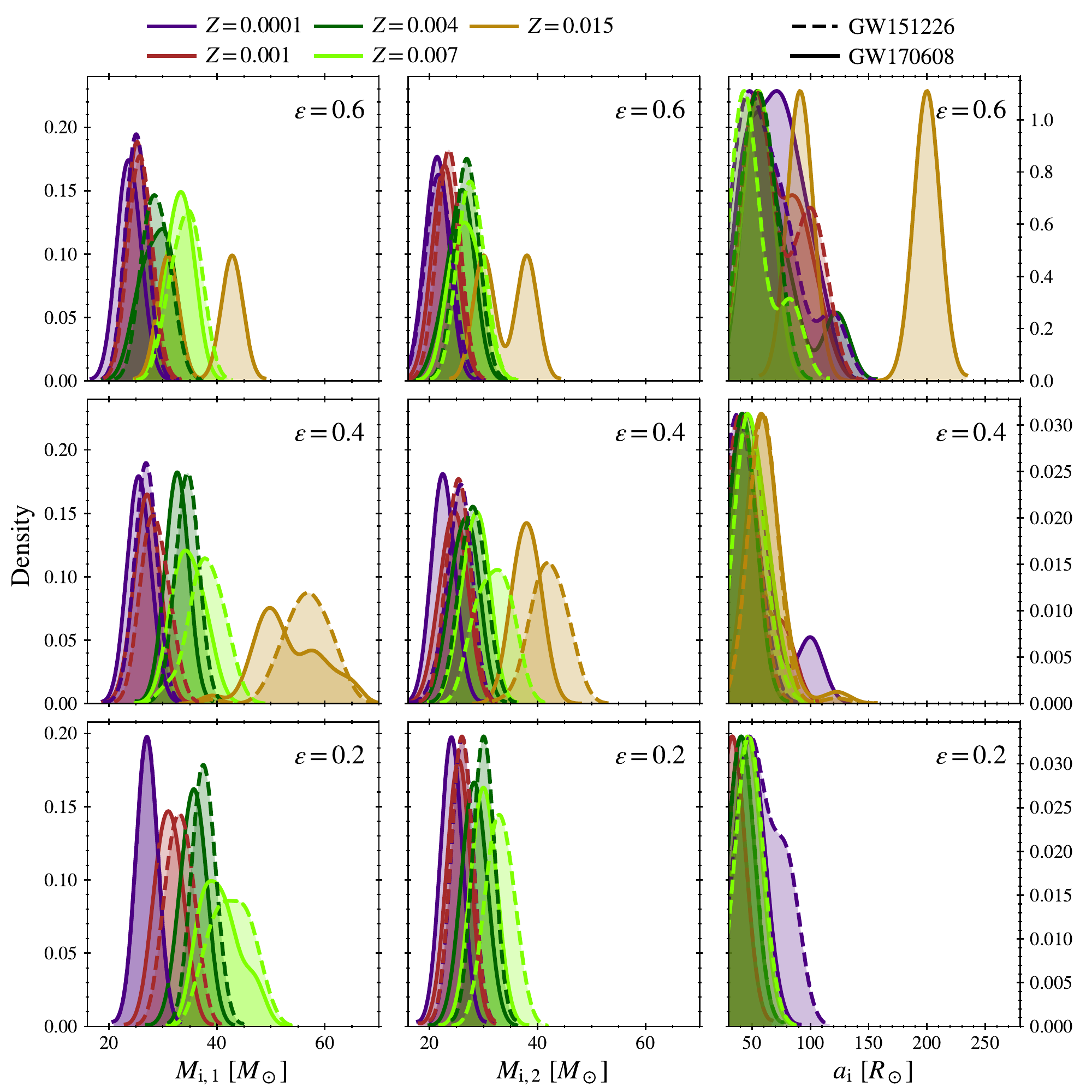}
    \caption{Population-weighted probability distributions for the parameters of the initial star binaries that eventually evolve in BBHs compatible with GW151226 (dashed) and GW170608 (solid) assuming $\alpha_{\rm CE}=2.0$. Left and middle panels show the component masses $M_{\rm i,1}$ and $M_{\rm i,2}$ of the initial binary and its initial separation $a_{\rm i}$ on the right. From top to bottom, the panels correspond to different MT efficiencies. Colours correspond to the metallicities given in the legend.}
    \label{fig:bps_param_a2}
\end{figure*}

% alphaCE = 1.0 plots (NEW!)
\begin{figure*}
    \centering
    \includegraphics[width=0.9\textwidth]{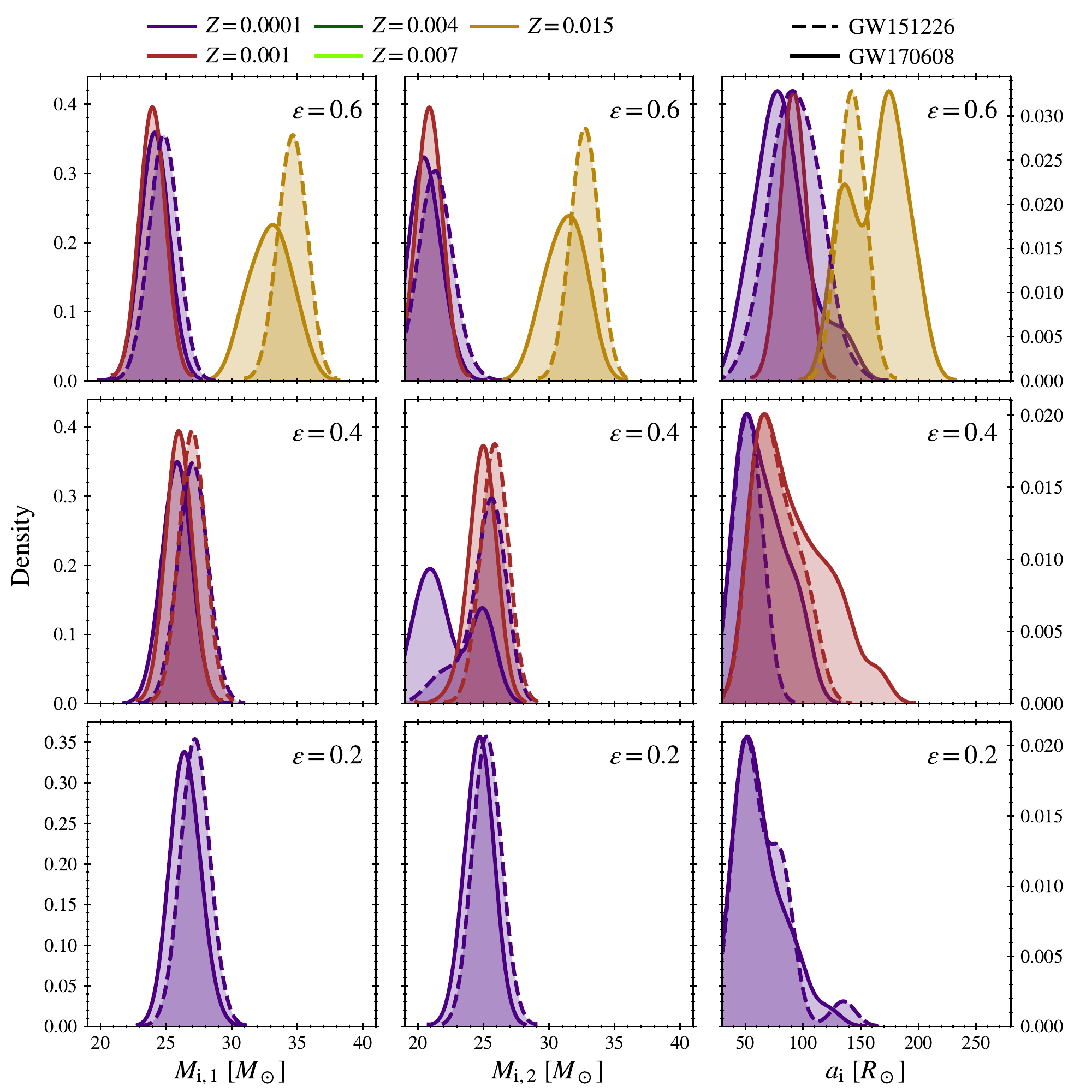}
    \caption{Idem to Figure~\ref{fig:bps_param_a2} for $\alpha_{\rm CE}=1.0$.}
    \label{fig:bps_param_a1}
\end{figure*}

Figures~\ref{fig:bps_param_a2}~and~\ref{fig:bps_param_a1} present the probability distributions for the parameters of the initial stellar binaries that eventually evolve into BBHs compatible with GW151226 (dashed lines) and GW170608 (solid lines), assuming $\alpha_{\rm CE}=2.0$ and $\alpha_{\rm CE}=1.0$, respectively. On the left and middle panels we show the initial masses of the progenitor binaries, $M_{\rm i,1}$ and $M_{\rm i,2}$, respectively. The panels on the right display the initial separations $a_{\rm i}$. From top to bottom, we show the results obtained with different MT efficiencies ($\epsilon$).

For $\alpha_{\rm CE}=2.0$ (Fig.~\ref{fig:bps_param_a2}), progenitors are found in the $\sim$20--40~M$_\odot$ range for the lower metallicities. For the higher metallicities, the initial masses move to $\sim$30--70~M$_\odot$ range, with a stronger dependence on the MT efficiency. In particular, progenitors with solar-like metallicity are not found in the low MT efficiency case (such as $\epsilon=0.2$, lowest panels). Initial separations cluster at values $\la$100~R$_\odot$ but solutions are found up to $\sim$250~R$_\odot$ for high metallicities. For $\alpha_{\rm CE}=1.0$ (Fig.~\ref{fig:bps_param_a1}), progenitors at solar-like metallicity ($Z=0.015$) are only found at the highest MT efficiencies $\epsilon=0.6$. For the lowest metallicity explored, $Z = 0.0001$, progenitors are found at every MT efficiency.

The initial masses of the binary progenitors have a clear dependence on metallicity, revealed by two aspects of the distributions: i) as metallicity increases, more massive progenitors (both primary and secondary masses) are required; ii) as metallicity increases, the initial-mass distributions widen. This can be interpreted as a consequence of the interplay between wind mass loss and initial binary separation. In initially wide binaries, stellar interactions occur later than in close binaries. Hence, the total mass loss due to stellar winds can operate on different time-scales depending on the initial separation. The higher the metallicity, the more pronounced this effect is. Progenitors masses also depend on the MT efficiency assumed. The more inefficient the MT process, the more massive progenitors are required to attain the proper target BH masses.

\subsubsection{Properties of the binary black holes}

\begin{figure}
    \centering
    \includegraphics[width=\hsize]{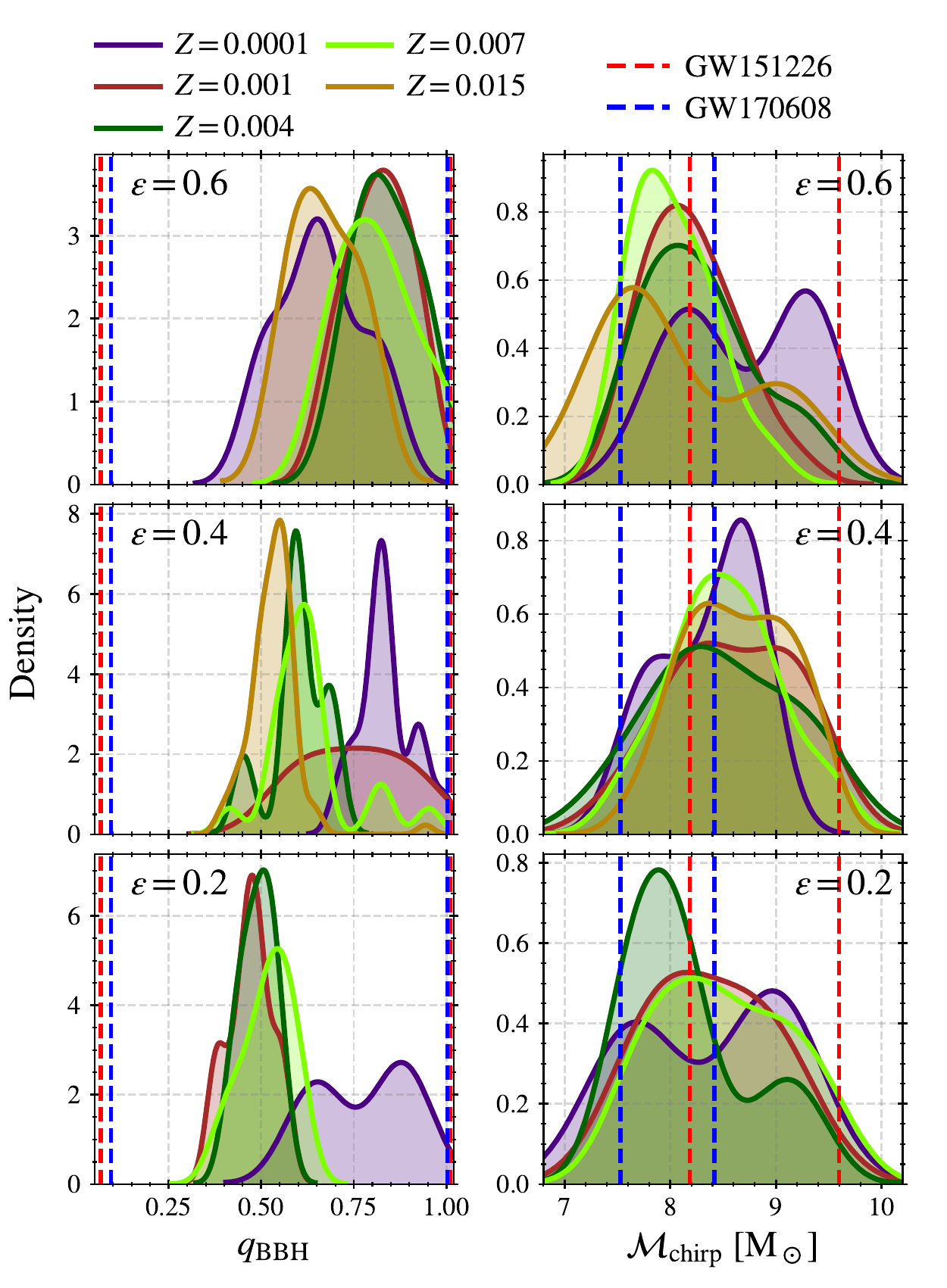}
    \caption{Population-weighted probability distributions for the parameter of the BBHs compatible with GW151226 and/or GW170608 assuming $\alpha_{\rm CE}=2.0$. Left and right panels correspond to mass ratio and chirp mass, respectively. From top to bottom, the panels correspond to different MT efficiencies.  Colours correspond to the metallicities given in the legend. The vertical dashed lines indicate the 100\% C.I. of $q_{\rm BBH}$ and $M_{\rm chirp}$ of GW151226 (red) and GW170608 (blue).}
    \label{fig:GWlikeKDEs_alpha2}
\end{figure}

\begin{figure}
    \centering
    \includegraphics[width=\hsize]{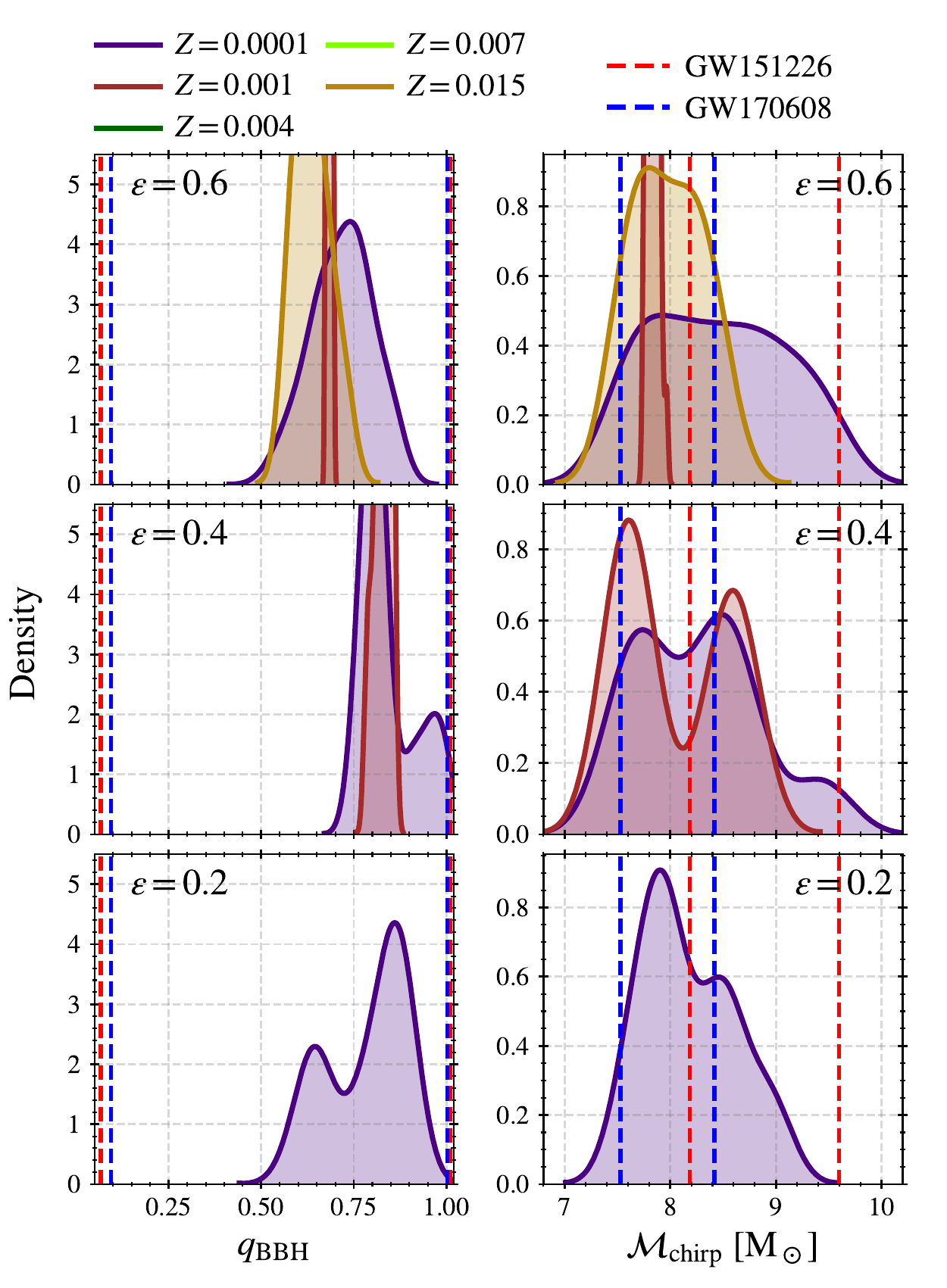}
    \caption{Same as Figure~\ref{fig:GWlikeKDEs_alpha2} for $\alpha_{\rm CE}=1.0$.}
    \label{fig:GWlikeKDEs_alpha1}
\end{figure}

Figures~\ref{fig:GWlikeKDEs_alpha2}~and~\ref{fig:GWlikeKDEs_alpha1} show the parameter distributions of the formed BBHs compatible with GW151226 and GW170608, assuming $\alpha_{\rm CE}=2.0$ and $\alpha_{\rm CE}=1.0$, respectively.

When $\alpha_{\rm CE}=2.0$ and $\epsilon \geq 0.4$, the smaller the metallicity the larger the mass of the secondary BH. When $Z$ decreases, the mass-ratio ($q_{\rm BBH} = M_{\rm{BH},2}/M_{\rm{BH},1}$) distribution peak shifts towards unity, and even exceeds $1$ for $Z \le 0.004$ and $\epsilon = 0.6$. For low MT efficiency ($\epsilon = 0.2$), secondary BHs are less massive, leading to $q_{\rm BBH} < 1$. For metallicities $Z \geq 0.001$, $q_{\rm BBH} \approx 0.5$. 

The chirp mass $\mathcal{M}_{\rm chirp}$ distribution basically spans the entire 100\% C.I. for both GW events, independently of the MT efficiency and metallicity. For the largest MT efficiency $\epsilon =0.6$, we note a slight preference to form less massive BBHs like GW170608 instead of GW151226.

When $\alpha_{\rm CE}=1.0$, the secondary BH is clearly the heaviest ($q_{\rm BBH} > 1$) when the MT efficiency is large, $\epsilon = 0.6$.
The chirp mass $\mathcal{M}_{\rm chirp}$ tends to decrease for the solar-like metallicity case. Several narrow distributions obtained are not fully reliable, due to the very low statistics available in this case, and only serve as a guide.

\subsubsection{Merger time delay}

\begin{figure}
    \centering
    \includegraphics[width=\hsize]{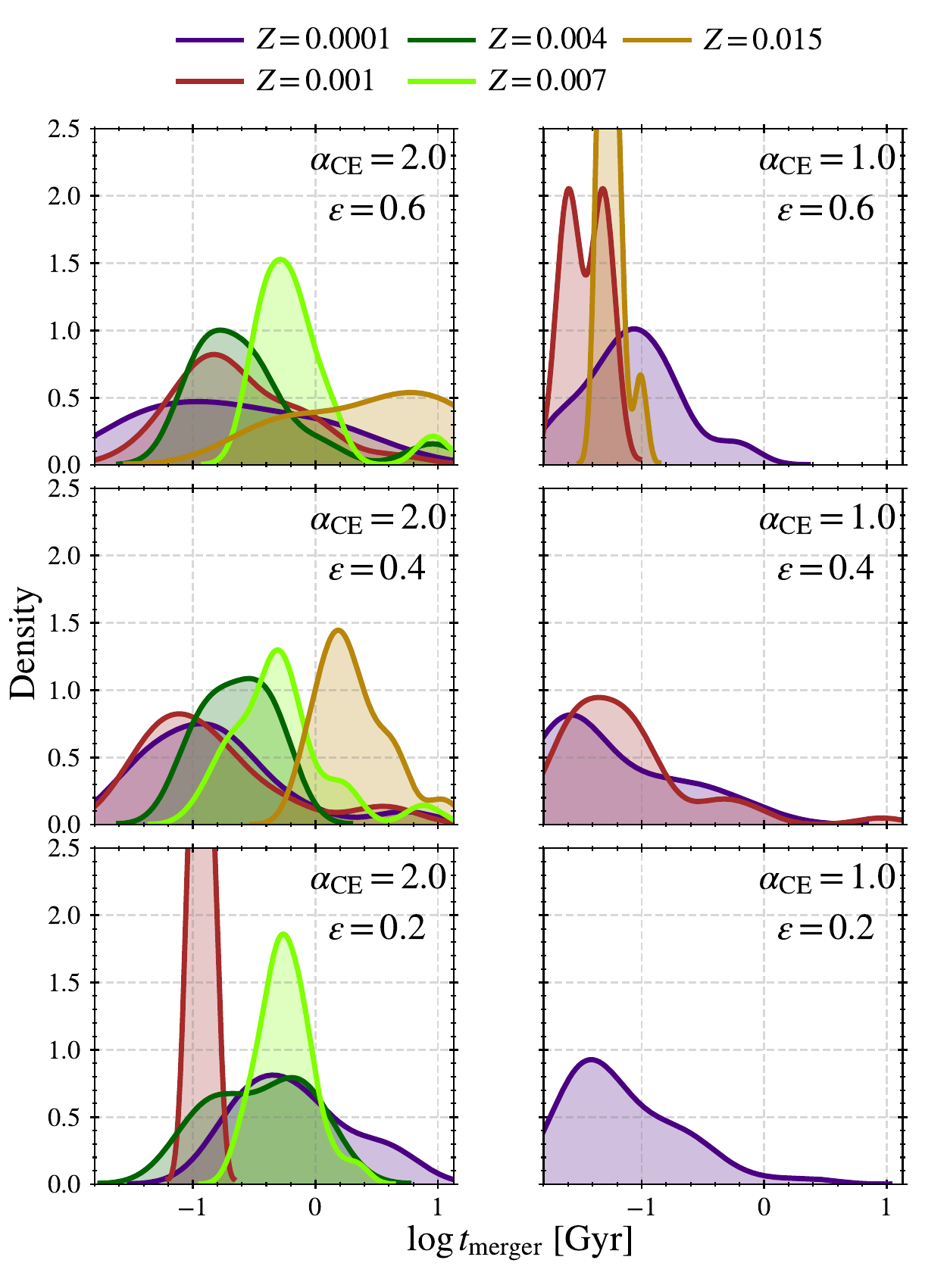}
    \caption{Population-weighted probability distribution of merger time delay ($t_{\rm merger}$) of BBHs compatible with GW151226 and/or GW170608 (within their 100\% C.I.). Left (right) panels correspond to $\alpha_{\rm CE}=2.0$ ($\alpha_{\rm CE}=1.0$). Top to bottom panels present different values of the MT efficiency adopted throughout this work. Different colours correspond to each metallicity value (see legend).} 
    \label{fig:merger_times_KDE}
\end{figure}

Fig.~\ref{fig:merger_times_KDE} presents the distribution of the merger time delay $t_{\rm merger}$. When $\alpha_{\rm CE} = 2.0$ (left panels), $t_{\rm merger}$ clearly increases with metallicity. The distribution peak goes from $\sim$100~Myr to $\sim$8~Gyr when $Z$ spans the selected metallicity range, from $0.0001$ to $0.015$. This correlation disappears for inefficient MT ($\epsilon = 0.2$) and, in this case, lower $t_{\rm merger}$ values are obtained in general.
When $\alpha_{\rm CE} = 1.0$, since final BBH are much more compact, merger time delays tend to be reduced by a factor $\sim 10$ with respect to the higher CE efficiency. The merger time delays are thus strongly impacted by the metallicity and the CE phase efficiency.

As shown in Fig.~\ref{fig:merger_times_KDE}, the merger time delays depend both on the CE efficiency and metallicity. The dependence on metallicity can be understood in terms of the angular momentum carried away by the stellar winds. As lower-metallicity binary progenitors lose less mass, their orbits do not experience significant widening. This leads to final smaller separations for lower-metallicity progenitors, compared to higher-metallicity ones. 

We notice that it is hard to make a thorough comparison of the merger delay time distributions with those found in population synthesis studies \citep{2012ApJ...759...52D,2018MNRAS.480.2011G}, given the low statistics inherent to our study, which is focused in a narrow range of $\mathcal{M}_{\rm chirp}$ and does not include natal kicks.

\section{Merger rate and gravitational-wave events}
\label{sec:merger-GW-event-rates}

We use the population-weighted samples presented in the previous section to estimate the local merger density rate leading to GW events comparable to those studied in this work.

\subsection{Method}

For binary distributions given by
%\begin{equation}
 $  {\rm d}N = f_{\rm j}(M_{\rm i,1}, M_{\rm i,2}, a_{\rm i}) \; {\rm d}x_{\rm j}$
%\end{equation}
our weighted simulations provide the number density of binaries in the multidimensional space defined by the initial masses, separations and delay times ($t_{\rm m} = T + t_{\rm merger} \approx t_{\rm merger}$, where $T \la 10$~Myr is the binary lifetime) which produce each specific GW event, defined as:
\begin{equation}
    \dfrac{{\rm d}N}{{\rm d}M_{\rm i,1} \, {\rm d}M_{\rm i,2} \, {\rm d}a_{\rm i} \, {\rm d}t_{\rm m}}(M_\mathrm{i,1},M_\mathrm{i,2},a_\mathrm{i}; t_\mathrm{m}) = P_{\rm GW-event} \; f_{\rm M_{\rm i,1}} \, f_{\rm M_{\rm i,2}} \, f_{\rm a_{\rm i}} 
\end{equation}
where $P_\mathrm{GW-event}$ is a Kronecker-delta function that selects binaries which evolve into BBHs compatible with the considered GW events.

By assuming a cosmology that relates the redshift $z$ to the cosmic time $t$, the intrinsic GW event rate $\mathcal{R}\left(Z,z(t)\right)$ can be obtained by integration over the full parameter space:
\begin{equation}\label{eq:rates}
    \begin{split}
        \mathcal{R}\left(Z,z(t)\right) = & \; \mathcal{N_\mathrm{corr}} \int_0^{t(z)} \int_{M_{\rm i,1}} \int_{M_{\rm i,2}} \int_{a_{\rm i}} \int_0^{t(z)} \dfrac{{\rm d}N}{{\rm d}M_{\rm i,1} \, {\rm d}M_{\rm i,2} \, {\rm d}a_{\rm i} \, {\rm d}t_{\rm m}} \\ 
        &  \widehat{{\rm SFR}} (t';Z) \delta \left[ t(z) - (t_{\rm m} + t') \right] {\rm d}t_{\rm m} {\rm d}a_{\rm i} \, {\rm d}M_{\rm i,2} \, {\rm d}M_{\rm i,1} \, {\rm d}t'
     \end{split}
\end{equation}
where $\mathcal{N_\mathrm{corr}}$ is a normalisation factor that includes the total mass $\mathcal{M}_{\rm T}$ of the $10^7$~simulated binaries, and takes into account the initial masses ($\mathcal{N}_{\rm IMF}$), mass ratios ($\mathcal{N}_q$) and separations ($\mathcal{N}_{a}$) excluded from the Monte-Carlo simulation
\begin{equation}\label{eq:corr-factors}
    \mathcal{N_{\rm corr}} = \dfrac{\mathcal{N}_a \mathcal{N}_{\rm IMF}}{\mathcal{N}_q (f_{\rm b})} \dfrac{1}{\mathcal{M}_{\rm T}};
\end{equation}
where we assume a binary fraction $f_{\rm b}=0.5$.

$\widehat{{\rm SFR}} (t';Z)$ is the metallicity-dependent star formation rate, namely
\begin{equation}
 \widehat{{\rm SFR}} (t';Z) = \mathrm{SFR}(t') \psi(Z,z'(t'))
\end{equation}
where $\mathrm{SFR}(t')$ is the total star formation rate history at binary-formation time $t'$ in co-moving coordinates \citep[which we adopt from][]{2004ApJ...613..200S}, and $\psi(Z,z'(t'))$ accounts for the fraction of stars formed at metallicity $Z$.

% Here is the equation from Strolger+ 2004:
%\begin{equation}
%    {\rm SFR}(t) = a \left( t^{b} \exp{(-t/c)} + d \exp{(d(t-t_{0})/c)} \right)    
%\end{equation}

We divide the full metallicity range into five intervals, namely $\Delta Z =$ 0--0.0005, 0.0005--0.0025, 0.0025--0.005, 0.005--0.0075, and 0.0075--0.03, which we assign to the five simulated values ($Z = 0.0001$, 0.001, 0.004, 0.007, and 0.015, see Sec.~\ref{sec:MESA}). We then compute $\Psi(\Delta Z,z')=\int_{\Delta Z} \psi(Z,z') \, {\rm d}Z$, where $\psi$ is normalised to unity, such that $\int_0^\infty \psi(Z,z') \, {\rm d}Z = 1$ at redshift $z'$ \citep{2006ApJ...638L..63L}.

Thanks to the $\delta$ function in Eq.~(\ref{eq:rates}), the summation runs over binary systems at redshift $z(t)$ with appropriate formation time $t'$ and merging delay times $t_{\rm m}$ and that evolve into BBH merging at cosmic time $t(z)$. In practice, this integral is evaluated by counting the fraction of sampled binaries per total simulated mass, $\mathcal{M}_{\rm T}$, that lead to a BBH merger at the expected redshift or cosmic time. The formation time and delay times are binned with a resolution of 100~Myr.

\subsection{Application}

\begin{table}
\caption{Merger rate density at zero redshift for each GW event and $\alpha_{\rm CE}$. Units are in yr$^{-1}$~Gpc$^{-3}$}  % title of Table
\label{table:ratesdensity_stats}      % is used to refer this table in the text
\centering                          % used for centering table
\begin{tabular}{c c c c c c}        % centered columns (4 columns)
\hline\hline                 % inserts double horizontal lines
{} & {} & \multicolumn{2}{c}{GW151226} & \multicolumn{2}{c}{GW170608} \\ % table heading 
\hline
$\epsilon$ & $Z$ & $\alpha_{\rm CE}=2$ & $\alpha_{\rm CE}=1$ & $\alpha_{\rm CE}=2$ & $\alpha_{\rm CE}=1$ \\    % table heading 
\hline                        % inserts single horizontal line
%0.6 & 0.015 & 0.202 & 0.004 & 0.347 & 0.050 \\
%    & 0.007 & 0.082 & 0.123 & 0.214 & 0.054 \\
%    & 0.004 & 0.143 & 0.025 & 0.186 & 0.005 \\
%    & 0.001 & 0.208 & 0.088 & 0.125 & 0.015 \\
0.6 & 0.015  & 0.032 & 0.069 & 0.529 & 0.308 \\
    & 0.007  & 0.975 & -- & 0.759 & -- \\
    & 0.004  & 0.575 & -- & 1.598 & -- \\
    & 0.001  & 0.362 & -- & 0.823 & 0.012 \\
    & 0.0001 & 0.117 & 0.018 & 0.063 & 0.017 \\
 & Total & 2.061 & 0.087 & 3.782 & 0.337 \\
\hline
%0.4 & 0.015 & 0.136 & -- & 0.366 & -- \\
%    & 0.007 & 0.138 & 0.028 & 0.136 & 0.019 \\
%    & 0.004 & 0.136 & -- & 0.213 & 0.018 \\
%    & 0.001 & 0.343 & 0.440 & 0.263 & 0.401 \\
0.4 & 0.015  & 3.603 & -- & 2.383 & -- \\
    & 0.007  & 1.116 & -- & 1.054 & -- \\
    & 0.004  & 0.183 & -- & 0.239 & -- \\
    & 0.001  & 0.344 & 0.087 & 0.265 & 0.190 \\
    & 0.0001 & 0.041 & 0.011 & 0.039 & 0.010 \\
 & Total & 5.287 & 0.098 & 3.980 & 0.200 \\
\hline
%0.2 & 0.015 & -- & -- & 0.009 & -- \\
%    & 0.007 & 0.121 & -- & 0.148 & -- \\
%    & 0.004 & 0.102 & -- & 0.116 & -- \\
%    & 0.001 & 0.043 & -- & 0.030 & -- \\
0.2 & 0.015  & -- & -- & -- & -- \\
    & 0.007  & 0.497 & -- & 0.464 & -- \\
    & 0.004  & 0.110 & -- & 0.323 & -- \\
    & 0.001  & 0.091 & -- & 0.125 & -- \\
    & 0.0001 & 0.024 & 0.008 & 0.017 & 0.013 \\
 & Total & 0.722 & 0.008 & 0.929 & 0.013 \\
%\hline                                   %inserts single line
%0.0 & 0.015 & -- & -- & -- & -- \\
%    & 0.007 & -- & -- & 0.004 & -- \\
%    & 0.004 & 0.007 & -- & 0.017 & -- \\
%    & 0.001 & 0.007 & -- & 0.009 & -- \\
\hline                                   %inserts single line
\end{tabular}

\end{table}

\begin{figure}
    \centering
    \includegraphics[width=\hsize]{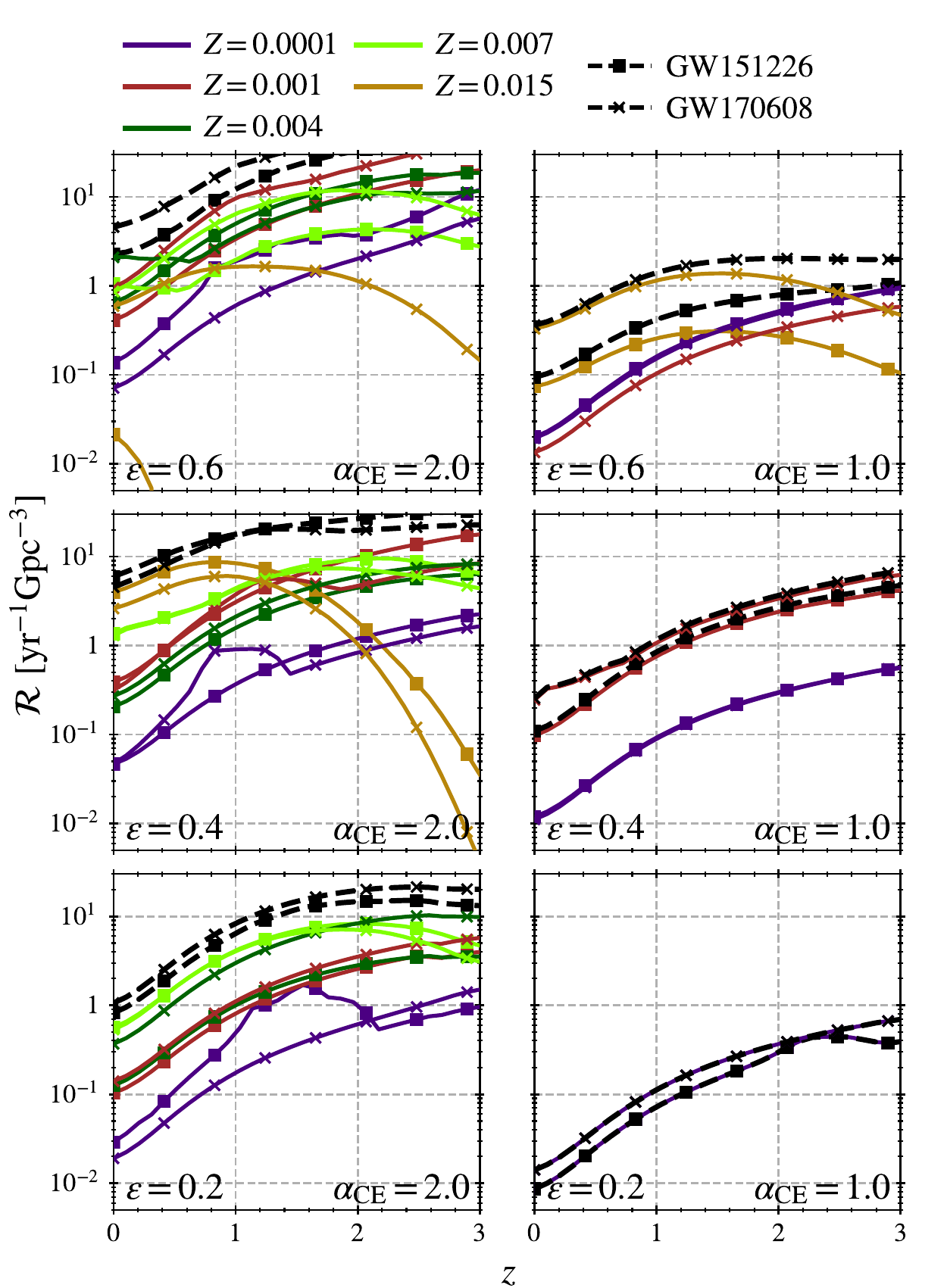}
    \caption{Merger rate density history of events compatible with GW151226 (cross markers) and GW170608 (square markers) as a function of redshift for each metallicity value adopted in this work (see legend for colours). Left panel (right panel) corresponds to simulations performed using $\alpha_{\rm CE}=2.0$ ($\alpha_{\rm CE}=1.0$). From top to bottom panels we show results for different MT efficiencies studied in this work.} 
    \label{fig:rates_on_metallicity}
\end{figure}

\begin{figure}
    \centering
    \includegraphics[width=\hsize]{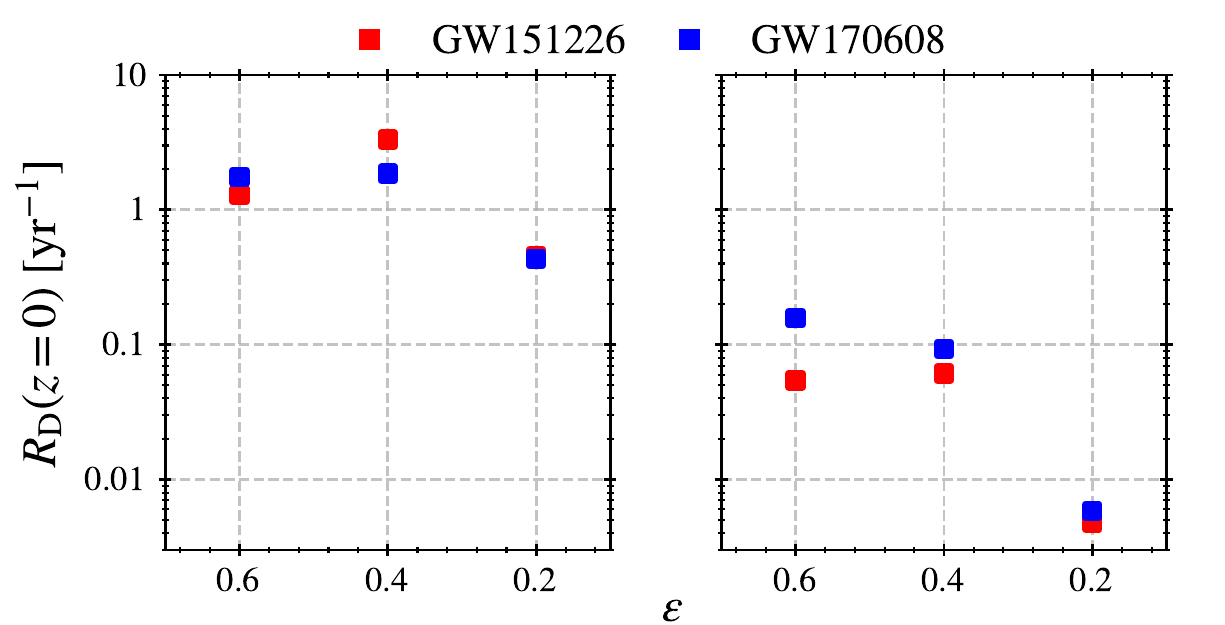}
    \caption{Total detection rates ($R_{D}(z=0)$) for O1+O2 LVC observing runs, marginalised over metallicity, as a function of MT efficiency $\epsilon$ for $\alpha_{\rm CE}=2.0$ (left panel) and $\alpha_{\rm CE}=1.0$ (right panel) of events compatible with GW151226 (red) and GW170608 (blue) within their 100\% credible intervals.} %Filled squares are detection rates of O1 and O2 LVC observing runs.}
    \label{fig:rates}
\end{figure}

Figure~\ref{fig:rates_on_metallicity} shows the dependency of the merger rate density $\mathcal{R}$ with the metallicity of the progenitor population and
Table~\ref{table:ratesdensity_stats} compares the local merger rate densities $\mathcal{R}(z=0)$, that are relevant for predicting GW event rates.

The expected local merger rate densities are larger for $\alpha_{\rm CE}=2.0$ in every case. This is consistent with the volume of the target regions in the parameter space of compatible binary progenitors, shown in Figs~\ref{fig:parameter-space-a2} and \ref{fig:parameter-space-a1}. As a direct consequence of the chemical evolution, the merger rates decay rapidly at high redshift for the largest metallicities ($Z = 0.015$~and 0.007), independently of the MT ($\epsilon$) and CE ($\alpha_{\rm CE}$) efficiencies. Moreover, at the present age ($z \approx 0$), the local merger rates decay, as a natural consequence of the decay in the SFR.

When $\alpha_{\rm CE} = 2.0$, in the local Universe, the rates are correlated with metallicity: the larger the metallicity, the larger the local merger rate density. This is more evident when $\epsilon = 0.4$, and less clear for $\epsilon=0.6$, where the contributions from all metallicities are more comparable between each other. A slight exception is found for $\epsilon = 0.2$, where no progenitors are found at the maximum (solar-like) metallicity explored. Moreover, in this latter case, the rates are significantly lower than in the former ones.

In the case of $\alpha_{\rm CE} = 1.0$, the rates decrease by an order of magnitude. The local merger rate density, $\mathcal{R}(z=0)$, is largely dominated by the lowest metallicities, except for $\epsilon=0.6$, where the high-metallicity progenitors dominate the rates.

\subsection{Implications for O1 and O2 science runs}

\begin{table}
\caption{Detection rates for O1+O2 LVC observing runs calculated using relation from \citet{2015ApJ...806..263D} for each GW event and $\alpha_{\rm CE}$, and considering $D_{\rm h} = 100$~Mpc.
} % title of Table
\label{table:rates_stats_o12}      % is used to refer this table in the text
\centering                          % used for centering table
\begin{tabular}{c c c c c}        % centered columns (4 columns)
\hline\hline                 % inserts double horizontal lines
{} & \multicolumn{2}{c}{GW151226} & \multicolumn{2}{c}{GW170608} \\ % table heading 
\hline
$\epsilon$ & $\alpha_{\rm CE}=2$ & $\alpha_{\rm CE}=1$ & $\alpha_{\rm CE}=2$ & $\alpha_{\rm CE}=1$ \\ % table heading 
\hline                        % inserts single horizontal line
%0.6 & $0.398$~yr$^{-1}$ & $0.151$~yr$^{-1}$ & $0.406$~yr$^{-1}$ & $0.058$~yr$^{-1}$ \\
%0.4 & $0.473$~yr$^{-1}$ & $0.292$~yr$^{-1}$ & $0.455$~yr$^{-1}$ & $0.204$~yr$^{-1}$ \\
%0.2 & $0.167$~yr$^{-1}$ & -- & $0.141$~yr$^{-1}$ & -- \\
%0.0 & $0.009$~yr$^{-1}$ & -- & $0.014$~yr$^{-1}$ & -- \\
0.6 & $1.293$yr$^{-1}$ & $0.054$yr$^{-1}$ & $1.757$yr$^{-1}$ & $0.157$yr$^{-1}$ \\
0.4 & $3.318$yr$^{-1}$ & $0.061$yr$^{-1}$ & $1.854$yr$^{-1}$ & $0.093$yr$^{-1}$ \\
0.2 & $0.453$yr$^{-1}$ & $0.005$yr$^{-1}$ & $0.432$yr$^{-1}$ & $0.006$yr$^{-1}$ \\
\hline                                   %inserts single line
\end{tabular}
\end{table}

We apply the relation from \citet{2015ApJ...806..263D} to rescale the intrinsic merger rate from \eqref{eq:rates} into GW detection rates:
\begin{equation}
 R_\mathrm{D} = \dfrac{4\pi}{3} D_\mathrm{h}^3 \bigl< w^3 \bigr> \bigl< 
                (\mathcal{M_\mathrm{c}}\,/\,1.2\,M_\odot)^{15/6} \bigr> \, \mathcal{R}(z=0)
\end{equation}
where $w$ is a geometrical factor, $\mathcal{M_\mathrm{c}}$ is the chirp mass of the BBH, $D_\mathrm{h}$ is the horizon distance and $\mathcal{R}$ is the merger rate density evaluated at $z=0$.

Consistently with the highest range for BNS obtained by advanced LIGO and advanced Virgo during their previous science run O2, we consider a binary neutron star (BNS) range $D_{\rm h}=100$~Mpc averaged over all sky directions. 
The results are shown in Fig.~\ref{fig:rates} and Table~\ref{table:rates_stats_o12}. The highest rates are obtained for the highest MT efficiencies ($\epsilon = 0.4$ and 0.6) in both CE cases. For the lowest MT efficiency, the outcome rates are significantly smaller: a factor of 4--5 in the high CE efficiency case, and a factor of $\sim$10 for the low CE efficiency. Thus, in general, the highest MT efficiency cases are favoured.

In Appendix \ref{app:SFRs}, we explore the dependence of the event rates on the assumed star formation history. We find that the strongest differences in event rates are introduced by the metallicity distribution. On the contrary, different SFR histories produce similar outcome rates. These results are compatible with those from \citet{2019MNRAS.488.5300C}, \citet{2019MNRAS.490.3740N} and \citet{2020MNRAS.493L...6T}.

\section{Discussion}
\label{sec:discussion}

In this work we have studied the progenitor properties for the two least-massive BBH mergers (GW151226 and GW170608) detected during the first two science runs of Advanced LIGO and Advanced Virgo, assuming they formed through the so-called isolated binary evolution channel. We simulated a large set of non-rotating stellar models using the binary stellar evolution code {\tt MESA} (see Appendix~\ref{app:mesa_runs}). We investigated a wide range of initial stellar masses, separations and metallicities. Moreover, to analyse the impact of unconstrained phases of binary evolution related to stellar interactions, we examined the dependence of the outcome results on different efficiencies for stable MT and CE ejection.

In the high CE efficiency scenario ($\alpha_{\rm CE}=2.0$), we found progenitors leading to BBH compatible with both GW events, for MT efficiencies $\epsilon \geq 0.2$. Their initial masses lay in the $20$--$65$~M$_\odot$ mass range for the primary (more massive) star and in the $18$--$48$~M$_\odot$ range for the secondary star. The initial separations are bound to the region $36$--$200$~R$_\odot$. The initial mass ranges depend strongly on the stellar metallicity. This is a direct consequence of the stellar wind efficiencies as pointed out by other authors \citep[e.g.,][]{2018MNRAS.480.2011G, 2018MNRAS.481.1908K}.
%, see Fig.~\ref{fig:parameter-space-a2}: the higher the metallicity, the more massive the progenitors need to be.

The results obtained in high CE efficiency regime are consistent with other studies in the literature based on different approaches. At low metallicity $Z=0.001$, our results are consistent with Fig.~1 from \citet{2017NatCo...814906S}. Furthermore, we obtain similar ranges for the progenitor masses for $Z = 0.004-0.007$ as \citet{2018MNRAS.481.1908K}. Although our highest metallicity differs, our progenitors for GW151226 consistently fall in rather lower mass ranges ($M_{{\rm i}, 1} \sim 45-65$~M$_\odot$ instead of $\sim$80~M$_\odot$, and $35 \lesssim M_{{\rm i}, 2} \lesssim 48$~M$_\odot$ instead of $55 \lesssim M_{{\rm i}, 2} \lesssim 60$~M$_\odot$) given the different BH formation scenario (it is worth mentioning that different MT and CE efficiencies were used).

In the low CE efficiency regime ($\alpha_{\rm CE}=1.0$) we obtain a narrower range of initial masses, favouring cases where initial masses are close to equal ($q \sim 1$), while initial separations tend to be shifted to higher values, see Fig.~\ref{fig:parameter-space-a1}. In this case a clear relation between the progenitor masses and the metallicity of the population is also recovered. We obtain solutions for the lowest metallicities explored ($Z=0.0001$ and $0.001$), which span largely on initial separations, thanks to the highly-suppressed mass-loss due to stellar winds, that lead to very stable target regions for the binary progenitors.

We find that all these binary systems undergo a CE phase when the primary star has already collapsed to a BH, as expected in the standard BBH formation scenario \citep{2002ApJ...572..407B, 2003MNRAS.342.1169V, 2006csxs.book..623T, 2012ApJ...759...52D, 2016Natur.534..512B}, with the companion star either crossing the Hertzsprung gap (HG) or already burning He in its core (CHeB). Although CE phases triggered while the donor star is in the HG when the star does not have a well-defined core-envelope structure \citep{2004ApJ...601.1058I} are usually assumed to lead to a CE merger \citep{2012ApJ...759...52D, 2019MNRAS.485..889S}, by means of our MT treatment (still 1D numerical simulations) we find regions of the explored parameter space populated with binary systems that survive such phase. The fraction of binaries in which the donor star is crossing the HG during the CE phase strongly depends on the metallicity due to its impact on the radial expansion of a star \citep{2020A&A...638A..55K}. We obtain that this fraction increases from $\sim 10$\% for $Z = 0.0001$ up to $\gtrsim$90\% for $Z = 0.015$. Moreover, the fraction of such binaries which finish the CE phase without a merger also increases with metallicity from $\sim$15\% to $\gtrsim$50\% from $Z = 0.0001$ to $Z = 0.015$, thus having a non-negligible contribution to the population of BBHs, especially at high metallicity. Our simulations show that these low-mass BBHs can only merge in timescales smaller than the Hubble time if they experience a CE phase enabling the ultra-compact binary formation (see also Appendix~\ref{app:time_delay}).

Additionally, assuming appropriate distributions for the initial-mass function, binary mass ratios and separations, we calculated the merger rates associated to each GW event for the explored metallicities, which could arise from different formation environments. In the case of $\alpha_{\rm CE}=2.0$, we find a correlation between the local merger rate and the metallicity: in general, the higher the metallicity, the larger the rate. This is more evident for $\epsilon = 0.4$. For the lowest MT efficiency, no progenitors are found at solar-like metallicity, which leads to a suppressed final rate. In this case, intermediate metallicities dominate. On the other hand, for $\alpha_{\rm CE}=1.0$ progenitors tend to be found at the lowest metallicities. We find that a decrease in the CE ejection efficiency produces lower rates in every case. The merger rate density history traces the SFR and thus the local merger rates peak at high redshift ($z \ga 1-2$). Moreover, the metallicity history has a strong impact on the local merger rates, due to younger solar-like metallicity progenitors with relatively short merger delay times.

In \citet{2015ApJ...814...58D} the authors show the impact of considering initial binary distributions taken from \citet{2012Sci...337..444S}, as the ones adopted in this paper, when compared to those from \citet{2012ApJ...759...52D}, which come from \citet{1983ARA&A..21..343A}. According to their Figures~1 and~2, the outcome progenitors of the full BBH population found using the former shows shorter periods than those found using the latter distribution. These progenitors mainly accommodate between 100-1000 days, similar to the progenitors that we find for the particular low-mass BBH population.

Care must be taken when comparing our derived rates with other works because our focus is set on the detection rate of two particular GW events. For the full BBH population, the LVC reported an empirical rate of $\mathcal{R} \simeq 9.7 - 101$~yr$^{-1}$~Gpc$^{-3}$ \citep{2019PhRvX...9c1040A} assuming a fixed population distribution, and a BBH merger rate density of $\mathcal{R} \simeq 53.2_{-28.2}^{+55.8}$~yr$^{-1}$~Gpc$^{-3}$ \citep{2019ApJ...882L..24A} using different models of the BBH mass and spin distributions, which are naturally higher than the values reported in this work. Our derived detection rates at instrumental sensitivity of Advanced~LIGO-Virgo detectors are $\sim$0.5--3~events per year for $\alpha_{\rm CE}=2.0$, and $\sim0.01-0.1$ for $\alpha_{\rm CE}=1.0$, with the former fully consistent with the actual GW events \citep[$\sim$2.1~yr$^{-1}$ for each event, considering one detection for a total of 167~d of coincident data for O1 and O2 runs, see][]{2016ApJ...832L..21A,2019PhRvX...9c1040A}.
In our simulations with $\alpha_{\rm CE}=2.0$ and for the highest MT efficiencies, we obtain rates which are consistent with those found by \citet{2018MNRAS.481.1908K}. However, in such cases we also find a comparable rates at intermediate metallicities. 

For our high-efficient CE ejection scenario, the lowest metallicities are disfavoured as progenitors of the observed low-mass GW events in the local Universe. % which prevent us from distinguishing the progenitor according to the metal content of the formation environment. 
In turn, in our simulations, the low-efficient CE scenario is highly disfavoured. In these cases, only low-metallicity progenitors are expected (except for the highest MT efficiency case) with very low merger delay times, which, combined with the metallicity history of the Universe, lead to local merger rates reduced at least by an order of magnitude.
Following this trend, we expect even smaller rates for lower CE efficiencies. Although we can not discard a non-negligible rate for $\alpha_{\rm CE}<1.0$, we focused on the region of the parameter space producing the largest expected rates, and compatible with the observed ones. Nevertheless, several recent population synthesis works focusing on the BBH population point to a high CE efficiencies ($\alpha_{\rm CE} > 1$) when the full BBH population is modelled \citep{2018MNRAS.480.2011G,2020ApJ...898..152S,2020arXiv201103564W}.

Finally, we caution that all these rates are subject to several uncertainties: when using different values in the input physical parameters, rates can vary by an order of magnitude. For example, it might be unlikely that the efficiency during MT phases remains the same throughout the entire evolution, as rotation might limit accretion from the companion \citep{1981A&A...102...17P, 1991ApJ...370..597P, 1991ApJ...370..604P}. Moreover, uncertainties in the mass-loss rates during the luminous blue variable and Wolf-Rayet phases could have an impact on the rates \citep{2018MNRAS.477.4685B}. Furthermore, metallicity evolution and star formation rate history were shown to have a strong impact on the BBH merger rates \citep[see, for instance,][and our Appendix \ref{app:SFRs}]{2019MNRAS.490.3740N}. In addition, including asymmetric kicks would also have an influence on the inferred rates. Since the nature of asymmetric kicks remains unknown, kicks are usually treated in a stochastic way. Including asymmetric kicks in our scheme would require running thousands of additional numerical simulations which fall out of the scope of this paper. Thus, in order to estimate the impact that asymmetric kicks could have on our results, in Appendix~\ref{app:kicks} we show the outcome of such simulations for a particular binary, leading in this case to a decrease in the intrinsic rates by a factor of $\sim$3 only.

\section{Summary and conclusions}
\label{sec:summary-conclusions}

We performed more than 60\,000 simulations of binary evolution with the 1D-hydrodynamic {\tt MESA} code, to study the formation history, progenitor properties and expected rates of the two lowest-mass BBH mergers detected during the O1 and O2 campaigns of LVC. To compute the whole evolution of the binary, we included {\it i.} the BH formation, through an instantaneous, spherically symmetric ejection, according to the {\em delayed} core-collapse prescription from \citet{2012ApJ...749...91F}; and {\it ii.} a numerical approach to simulate the CE phase (with two values of the efficiency parameter $\alpha_{CE} = 1.0$ and 2.0).

Our modelling contains simplified assumptions and limitations that are worth
to enumerate in this summary. i) Asymmetric kicks during BH formation are not incorporated (but see Appendix~\ref{app:kicks} for a discussion on the impact expected from natal kicks); ii) BH accretion during CE phase is considered negligible. This effect could lead to slightly higher BH masses and consequently, less massive progenitors (but see a discussion in Sections~\ref{sec:MESA}~and~\ref{sec:merger-GW-event-rates} about the theoretical uncertainties on this particular subject); iii) $\alpha_{\rm CE} < 1$ is not explored, based on the inferred rates obtained for $\alpha_{\rm CE} = 1.0$ and 2.0; iv) initially eccentric binaries are not considered mainly due to computational limitations; orbits may circularise even before the MT onset, or on a short timescale during the first MT episode \citep{1995A&A...296..709V}. This limitation will have an impact on the distribution of initial binary separations that lead to the GW events under study. v) The effects of rotation and tides on the internal mixing are not taken into account.

We summarise below the main results achieved in this work:

\begin{enumerate}
    \item {\bf General remarks:} the stellar progenitors of GW\,151226 are more massive than those of GW\,170608 (in agreement with the final masses of the black holes); higher initial orbital separation $a_i$ implies longer merger times $t_{\rm merger}$; higher metallicity $Z$ implies more massive progenitors (due to mass lost through stellar winds); no progenitors are found for the fully inefficient mass transfer MT ($\epsilon=0$); for the low-efficiency MT case ($\epsilon = 0.2$), only low $Z \le 0.001$ binaries can become progenitors, and for high MT efficiency ($\epsilon \geq 0.4$), we obtain either solar-like Z progenitors of different masses, or low Z progenitors evolving towards similar mass stars (mass ratio $q$ close to unity, due to rejuvenation process, where the second-formed BH becomes more massive than --or at least as massive as-- the first); In the case of low CE efficiency ($\alpha_{\rm{CE}}=1.0$), we obtain progenitors having $q$ close to unity (rejuvenation), having only low $Z=0.001-0.0001$, except for the highest MT efficiency, where also solar-like Z progenitors are found.

    \item {\bf Mass ratio and chirp masses:} $q_{\rm BBH}$ is always > 0.4, covering all the $M_{\rm chirp}$ range; high MT efficiencies ($\epsilon \geq 0.4$) tend to form BBH at any $q_{BBH}$, while $q_{BBH} \sim 0.4-0.6$ for $\epsilon=0.2$. Low Z stars span whole range of $q_{BBH}$, showing decreasing $M_{\rm chirp}$ as $q_{\rm BBH}$ increases. Rejuvenated stars at the highest MT efficiencies lead to $q_{\rm BBH} \sim 1$. For $\alpha_{\rm CE}=1.0$, progenitors tend towards equal-mass binaries, with all BBHs having $q_{\rm BBH} > 0.6$ at all Z, and even $q_{\rm BBH} > 1.0$ for $\epsilon = 0.6$ (rejuvenation process).
    
    \item {\bf Merger time delay:} for $\alpha_{\rm{CE}}=2.0$, $t_{\rm merger}$ increases with metallicity, from 10~Myr to 10~Gyr (no correlation though for $\epsilon = 0.2$, for which $t_{\rm merger} \sim 0.1-2$\,Gyr), while for $\alpha_{\rm CE}=1.0$, $t_{\rm merger}$ is much shorter (due to late ejection of CE), from $\sim$5~Myr to $\la$1~Gyr (typically 100\,Myr); There exists a dichotomy between an old merger population made of high Z progenitors, and a young merger population constituted of low Z progenitors; The merger time delay is strongly impacted by both the metallicity and the assumed CE efficiency, the CE phase being always required for binaries to merge within the Hubble time.
    
    \item {\bf Merger rate density:} Local merger rate densities $\mathcal{R}(z=0)$ are all larger for $\alpha_{\rm{CE}}=2.0$ than $\alpha_{\rm{CE}}=1.0$. %(only for $\epsilon = 0.4$ and $Z=0.001$ is $\mathcal{R}$ 40\% higher for $\alpha_{\rm{CE}}=1.0$ than for $\alpha_{\rm{CE}}=2.0$); 
    $\mathcal{R}$ decays rapidly at high redshift for large metallicity (due to chemical evolution of the universe), independently of $\alpha_{\rm{CE}}$. For $\alpha_{\rm{CE}} = 2.0$, $\mathcal{R}\ga1$ for $\epsilon \geq 0.4$; For $\alpha_{\rm{CE}}=1.0$, $\mathcal{R}$ is mainly dominated by low Z, independently of MT rate.
    
\end{enumerate}

As a future work we plan to extend the range of masses of the binary progenitors studied here in order to explore the low-mass end of BH formation, and its transition to neutron stars, which could lead to a mass gap in the compact object masses, that might be probed with GW observations of BBHs. In addition, more comprehensive modelling, including stellar rotation and asymmetric kicks, is also on the scope of future projects.

\begin{acknowledgements}
We are grateful to the Referee whose insightful comments helped us to improve the quality of this paper. This work was supported by the LabEx UnivEarthS, Interface project I10, ``From evolution of binaries to merging of compact objects''. We acknowledge use of Arago Cluster from Astroparticule et Cosmologie (APC) for our calculations. ASB is a CONICET fellow. We are grateful to the {\tt MESA} developers for building and making available high-quality computational software for astrophysics.\\

{\em Software:} {\tt MESA:} Modules for Experiments in Stellar Astrophysics\footnote{\url{http://mesa.sourceforge.net/}}, {\tt ipython/jupyter} \citep{2007CSE.....9c..21P}, {\tt matplotlib} \citep{2007CSE.....9...90H}, {\tt NumPy} \citep{2011CSE....13b..22V}, {\tt scipy} \citep{scipy} and {\tt Python} from \url{python.org}. This research made use of {\tt astropy}, a community-developed core {\sc Python} package for astronomy \citep{2013A&A...558A..33A,2018AJ....156..123A}.
\end{acknowledgements}

\bibliographystyle{aa}
\bibliography{aanda}

\begin{appendix} %appendix

\section{{\tt MESA} runs: full parameter space exploration}
\label{app:mesa_runs}

As explained in Section~\ref{sec:mesa-runs}, in the frame of this work we explored a wide range of the parameter space defined by the binary initial parameters: $M_{\rm i,1}$, $M_{\rm i,2}$ and $a_{\rm i}$ with the main goal of finding the target regions of solutions that correspond to models compatible with binary progenitors of the GW170608 and GW151226 events. This task was performed for four different values of MT efficiencies and metallicities. For this purpose, we started by the exploration of the target regions corresponding to $\alpha_{\rm{CE}}=2.0$, which naturally leads to a higher amount of solutions compatible with the GW events with respect to $\alpha_{\rm{CE}}=1.0$, since the fraction of CE mergers is much lower as the CE is more efficiently removed. For this we used a grid of even numbers for $M_{\rm i,1}$ and odd numbers for $M_{\rm i,2}$ ($\Delta M = 2$~M$_\odot$) and a logarithmic separation in $a_{\rm i}$ of 0.02 dex. We first started by simulating progenitor masses giving CO cores leading to BHs compatible with the observed BH masses and later expanding the regions until no compatible solutions were found. Once these target regions were fully covered, we switched to the exploration of the $\alpha_{\rm{CE}}=1.0$ case. For this, since we already counted with the initial exploration, we focused on the binary models that lead to CE triggers, which we re-run using the low CE efficiency. Since these target regions are naturally smaller, we decreased the grid to $\Delta M = 1$~M$_\odot$ and 0.01~dex for $a_{\rm i}$ to have a better coverage. For each CE survival we also simulated the neighbours in the grid until the target regions were fully covered, in an iterative fashion.

\begin{figure*}
   \centering
        \includegraphics[width=\hsize]{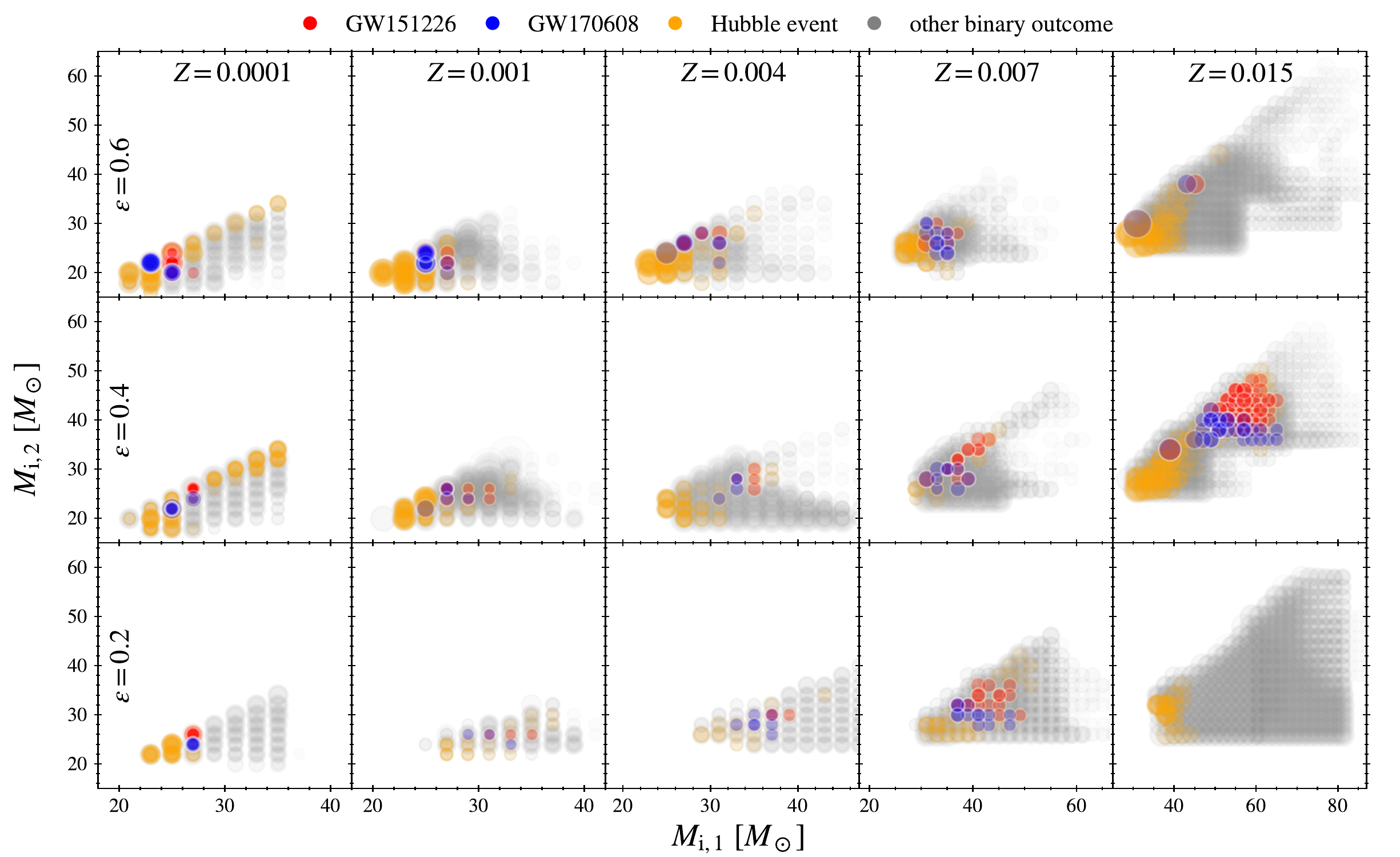}
            \caption{Full parameter space explored using $\alpha_{\rm{CE}}=2.0$. 
            Blue (red) circles show compatible models with GW170608 (GW151226). Orange circles represent models leading to BBH that merge within the Hubble time, while grey circles are used for the rest of the simulations.
            }
         \label{fig:appendix-runs-a2}
\end{figure*}

\begin{figure*}
   \centering
        \includegraphics[width=\hsize]{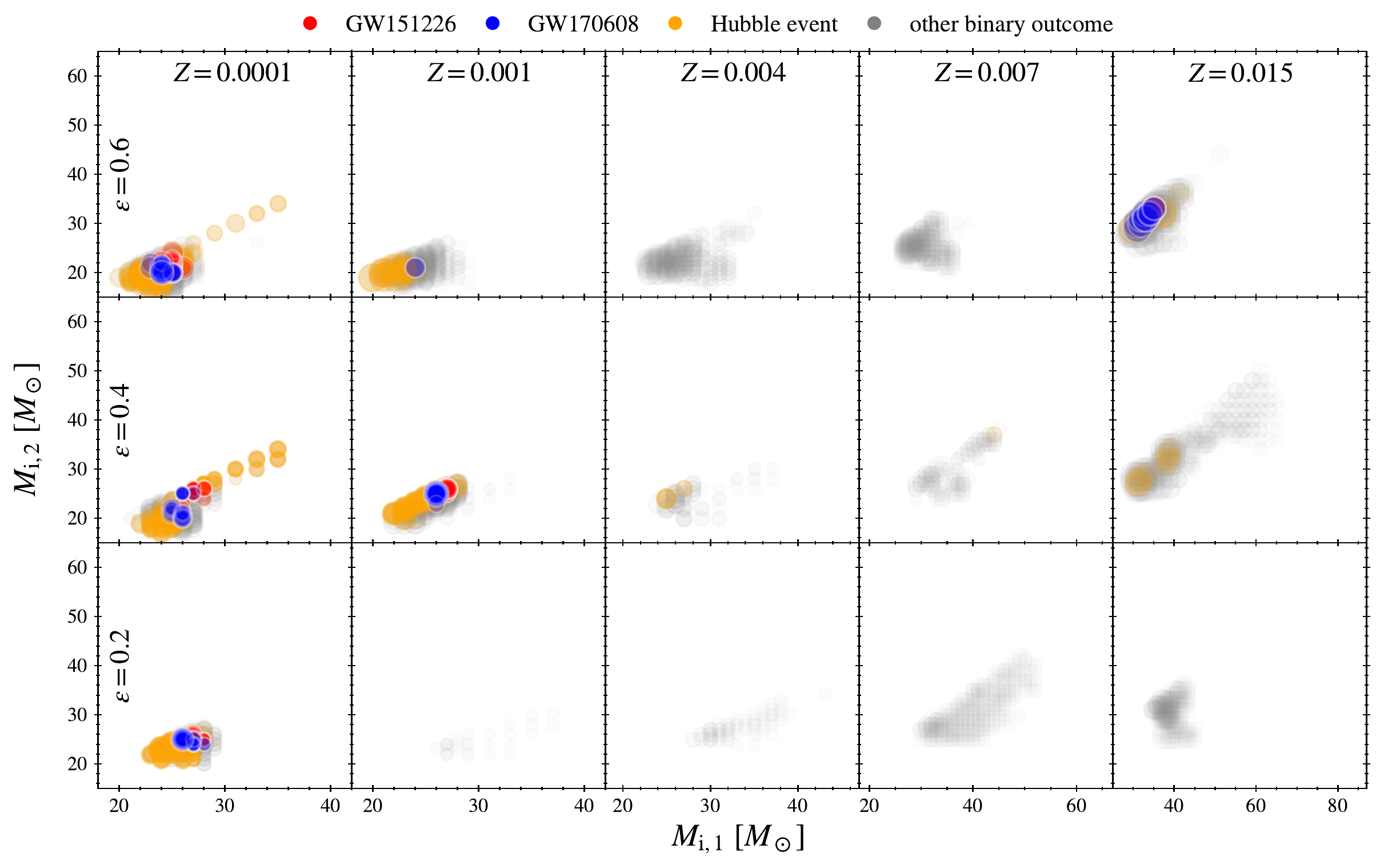}
            \caption{Idem to Figure~\ref{fig:appendix-runs-a2} for $\alpha_{\rm CE}=1.0$.}
         \label{fig:appendix-runs-a1}
\end{figure*}

In Figures~\ref{fig:appendix-runs-a2}~and~\ref{fig:appendix-runs-a1} we present the full parameter space explored using $\alpha_{\rm{CE}}=2.0$ and $1.0$, respectively. Panels from top to bottom correspond to each set of MT efficiencies: $\epsilon=0.6$, $0.4$ and $0.2$; we do not show the completely inefficient MT case as no compatible progenitors were found. Panels from left to right correspond to each set of metallicities: $0.0001$, $0.001$, $0.004$, $0.007$, and $0.015$. Blue (red) circles are used for models compatible with GW170608 (GW151226). The size of the circles is proportional to the initial separation ($a_i$). Orange circles represent models leading to BBHs that merge within the Hubble time but with $\mathcal{M}_{\rm chirp}$ incompatible with the GW events considered. Grey circles are used for the rest of the models used for this work. In Table~\ref{table:appendix_runs_a2}~and~\ref{table:appendix_runs_a1} we summarise the main characteristics of all the runs performed, including total runs, total of runs leading to BBHs, total runs leading to BBHs that merge within the Hubble time, total runs compatible with GW170608 and GW151226 and the ranges covered in the parameter space defined by $M_{\rm i,1}$, $M_{\rm i,2}$ and $a_i$ for each MT efficiency ($\epsilon$) and metallicity ($Z$).

\begin{table*}
\caption{Summary of {\tt MESA} runs performed with $\alpha_{\rm CE}=2.0$.}  % title of Table
\label{table:appendix_runs_a2}
\centering                 
\begin{tabular}{l l c c c c c c c c c }   
%\hline\hline
%$\alpha_{\rm CE}=2.0$ \\
\hline\hline
$\epsilon$ & $Z$ & Runs & BBH & BBH (<t$_{\rm Hubble}$) & GW170608 & GW151226 & $M_{\rm i,1}$ [M$_\odot$] & $M_{\rm i,2}$ [M$_\odot$] & $a_i$ [R$_\odot$] \\
\hline\hline
%0.6 & 0.001 & 1637 & 1042 & 261 & 31 & 48 & 23--45 & 14--36 & 36--300 \\
% & 0.004 & 1160 & 743 & 191 & 46 & 51 & 27--45 & 20--42 & 36--150 \\
% & 0.007 & 1820 & 1070 & 159 & 45 & 22 & 27--57 & 24--44 & 36--200 \\
% & 0.015 & 5207 & 2617 & 288 & 80 & 70 & 31--89 & 26--78 & 36--200 \\
0.6 & 0.0001 & 595 & 455 & 111 & 14 & 25 & 21--35 & 16--34 & 27--150 \\
    & 0.001 & 1727 & 436 & 120 & 16 & 8 & 21--45 & 14--36 & 30--300 \\
    & 0.004 & 1245 & 419 & 95 & 7 & 7 & 23--45 & 18--42 & 30--200 \\
    & 0.007 & 1854 & 503 & 60 & 11 & 5 & 27--57 & 20--44 & 36--200 \\
    & 0.015 & 5223 & 1520 & 81 & 2 & 1 & 29--89 & 26--78 & 36--200 \\
\hline
%0.4 & 0.001 & 1532 & 1253 & 232 & 42 & 47 & 27--45 & 20--36 & 30--500 \\
% & 0.004 & 2187 & 1403 & 205 & 33 & 41 & 27--53 & 20--48 & 36--200 \\
% & 0.007 & 2424 & 1382 & 266 & 53 & 71 & 31--65 & 24--58 & 36--200 \\
% & 0.015 & 3458 & 1621 & 450 & 114 & 51 & 31--85 & 26--78 & 52--200 \\
0.4 & 0.0001 & 636 & 497 & 115 & 7 & 10 & 21--37 & 18--34 & 27--186 \\
    & 0.001 & 1669 & 476 & 76 & 5 & 8 & 21--45 & 20--36 & 30--200 \\
    & 0.004 & 2223 & 697 & 54 & 4 & 4 & 25--53 & 20--48 & 33--200 \\
    & 0.007 & 2453 & 681 & 34 & 9 & 15 & 29--65 & 24--58 & 36--200 \\
    & 0.015 & 3549 & 1382 & 192 & 31 & 59 & 29--85 & 26--78 & 43--200 \\
\hline
%0.2 & 0.001 & 986 & 620 & 123 & 16 & 30 & 27--47 & 24--46 & 30--100 \\
% & 0.004 & 1202 & 750 & 251 & 41 & 66 & 29--61 & 24--42 & 36--91 \\
% & 0.007 & 2741 & 1496 & 430 & 89 & 138 & 31--83 & 26--50 & 36--150 \\
% & 0.015 & 4962 & 3077 & 318 & 1 & 0 & 37--91 & 26--58 & 36--150 \\
0.2 & 0.0001 & 481 & 349 & 51 & 5 & 6 & 23--39 & 20--34 & 30--122 \\
    & 0.001 & 1019 & 261 & 24 & 3 & 3 & 25--47 & 22--46 & 30--100 \\
    & 0.004 & 1221 & 424 & 21 & 7 & 3 & 29--61 & 24--42 & 36--91 \\
    & 0.007 & 2750 & 961 & 59 & 13 & 17 & 29--83 & 26--50 & 36--150 \\
    & 0.015 & 5019 & 2996 & 35 & 0 & 0 & 36--91 & 26--58 & 36--150 \\
\hline
%0.0 & 0.001 & 348 & 324 & 25 & 5 & 5 & 29--41 & 26--34 & 36--100 \\
% & 0.004 & 744 & 592 & 28 & 8 & 6 & 37--49 & 26--46 & 36--100 \\
% & 0.007 & 1102 & 766 & 13 & 3 & 0 & 43--57 & 30--54 & 36--150 \\
% & 0.015 & 2135 & 1876 & 0 & 0 & 0 & 43--79 & 32--56 & 30--2000 \\
0.0 & 0.0001 & 137 & 85 & 0 & 0 & 0 & 29--37 & 24--32 & 30--63 \\
    & 0.001 & 348 & 207 & 0 & 0 & 0 & 29--41 & 26--34 & 36--100 \\
    & 0.004 & 744 & 472 & 0 & 0 & 0 & 37--49 & 26--46 & 36--100 \\
    & 0.007 & 1102 & 554 & 0 & 0 & 0 & 43--57 & 30--54 & 36--150 \\
    & 0.015 & 2135 & 1405 & 0 & 0 & 0 & 43--79 & 32--56 & 30--2000 \\
\hline
\end{tabular}
\end{table*}

\begin{table*}
\caption{Summary of {\tt MESA} runs performed with $\alpha_{\rm CE}=1.0$.}  % title of Table
\label{table:appendix_runs_a1}
\centering                 

\begin{tabular}{l l c c c c c c c c c }   
%\hline\hline
%$\alpha_{\rm CE}=1.0$ \\
\hline\hline
$\epsilon$ & $Z$ & Runs & BBH & BBH (<t$_{\rm Hubble}$) & GW170608 & GW151226 & $M_{\rm i,1}$ [M$_\odot$] & $M_{\rm i,2}$ [M$_\odot$] & $a_i$ [R$_\odot$] \\
\hline\hline
%0.6 & 0.001 & 2043 & 1095 & 179 & 5 & 57 & 22--55 & 18--42 & 36--315 \\
% & 0.004 & 736 & 186 & 14 & 1 & 6 & 27--45 & 19--41 & 36--174 \\
% & 0.007 & 2084 & 994 & 65 & 12 & 36 & 27--55 & 23--43 & 43--180 \\
% & 0.015 & 1637 & 768 & 75 & 5 & 2 & 30--85 & 26--62 & 43--220 \\
0.6 & 0.0001 & 2106 & 482 & 241 & 35 & 42 & 20--45 & 16--36 & 30--190 \\
    & 0.001 & 2296 & 200 & 43 & 2 & 0 & 20--55 & 18--42 & 30--315 \\
    & 0.004 & 1007 & 115 & 0 & 0 & 0 & 23--45 & 18--41 & 33--220 \\
    & 0.007 & 2337 & 102 & 0 & 0 & 0 & 27--55 & 20--43 & 36--180 \\
    & 0.015 & 1822 & 194 & 20 & 11 & 3 & 29--85 & 26--62 & 43--220 \\
\hline
%0.4 & 0.001 & 2454 & 2093 & 986 & 205 & 337 & 26--45 & 20--37 & 33--314 \\
% & 0.004 & 422 & 211 & 30 & 4 & 1 & 26--53 & 20--44 & 40--105 \\
% & 0.007 & 1107 & 469 & 71 & 5 & 27 & 28--65 & 24--50 & 40--150 \\
% & 0.015 & 4713 & 1511 & 39 & 0 & 0 & 30--81 & 25--59 & 52--210 \\
0.4 & 0.0001 & 2552 & 586 & 228 & 23 & 27 & 21--45 & 17--36 & 30--135 \\
    & 0.001 & 2856 & 406 & 142 & 24 & 19 & 22--45 & 19--37 & 30--314 \\
    & 0.004 & 525 & 39 & 6 & 0 & 0 & 24--53 & 20--44 & 36--105 \\
    & 0.007 & 1319 & 69 & 1 & 0 & 0 & 28--65 & 24--50 & 36--150 \\
    & 0.015 & 5017 & 99 & 10 & 0 & 0 & 30--81 & 25--59 & 48--210 \\
\hline
%0.2 & 0.001 & 482 & 328 & 95 & 0 & 0 & 26--47 & 22--38 & 36--141 \\
% & 0.004 & 844 & 295 & 9 & 0 & 0 & 29--62 & 24--43 & 43--83 \\
% & 0.007 & 2255 & 620 & 2 & 0 & 0 & 31--79 & 25--51 & 43--122 \\
% & 0.015 & 852 & 328 & 0 & 0 & 0 & 40--68 & 25--55 & 43--142 \\
0.2 & 0.0001 & 3001 & 457 & 219 & 31 & 21 & 22--45 & 20--36 & 30--190 \\
    & 0.001 & 526 & 6 & 0 & 0 & 0 & 26--47 & 22--38 & 30--141 \\
    & 0.004 & 968 & 14 & 0 & 0 & 0 & 28--62 & 24--43 & 40--83 \\
    & 0.007 & 2923 & 32 & 0 & 0 & 0 & 30--79 & 25--51 & 40--122 \\
    & 0.015 & 1247 & 13 & 0 & 0 & 0 & 35--68 & 25--55 & 43--142   \\
%\hline
%0.0 & 0.001 & 50 & 2 & 0 & 0 & 0 & 29--39 & 26--34 & 36--63 \\
% & 0.004 & 28 & 1 & 0 & 0 & 0 & 37--47 & 26--38 & 40--57 \\
% & 0.007 & 13 & 1 & 0 & 0 & 0 & 43--53 & 30--36 & 43--63 \\
% & 0.015 & -- & -- & -- & -- & -- & -- & -- & -- \\
\hline
\end{tabular}
\end{table*}

\section{{\tt MESA} example}
\label{app:example}

In order to illustrate the evolutionary channel explored throughout this work, we present the full evolution of a binary system which ends its evolution as a BBH compatible with one of the events detected by the LVC, using the two values for the CE efficiency ($\alpha_{\rm CE}=2.0$ and 1). The initial parameters for the model are: $M_{{\rm i},1} = 35$~M$_\odot$, $M_{{\rm i},2} = 32$~M$_\odot$, $a_{\rm i} = 83.05$~R$_\odot$, $Z = 0.007$, and a MT efficiency of $\epsilon=0.4$.
 
\begin{figure}
    \centering
    \includegraphics[width=\hsize]{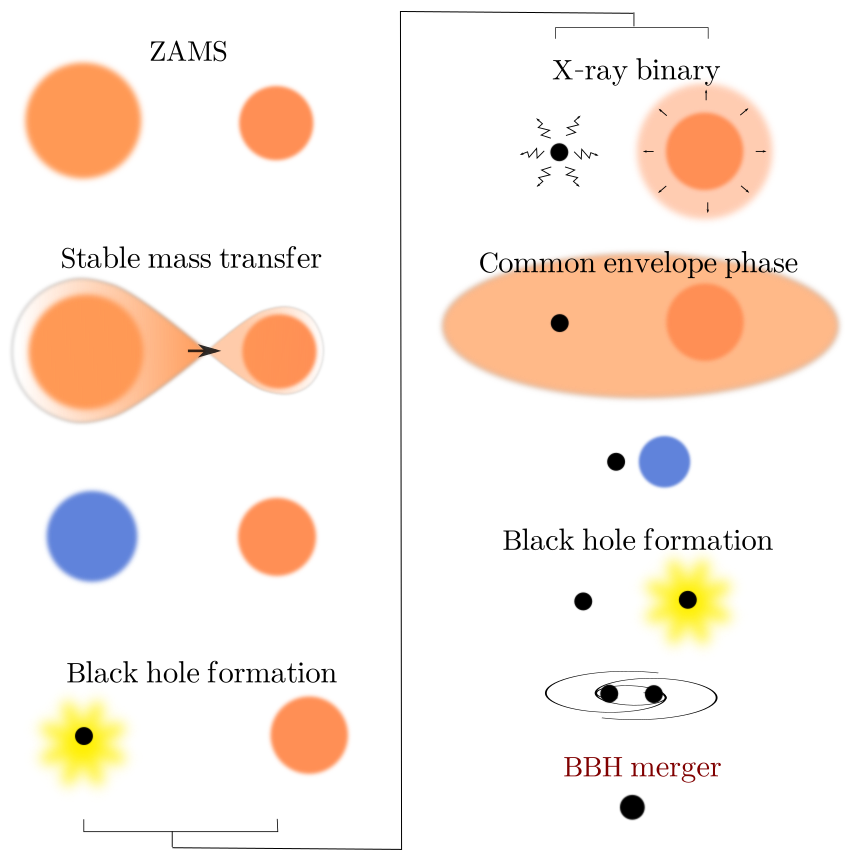}
    \caption{Schematic view of the binary evolutionary channel explored throughout this work.}
    \label{fig:bin2bbh}
\end{figure}

In Figure~\ref{fig:bin2bbh} we present a scheme for the full binary evolution channel followed by the system from ZAMS to the BBH merger. In general, once the primary (most massive) star expands, the system experiences an initial stable MT phase until the primary contracts and later collapses to form a first BH. Later on, once the secondary expands, a new phase of stable MT develops and the system becomes an X-ray binary. If the MT becomes unstable, a short CE phase is triggered, and the binary separation is heavily reduced while the donor envelope is removed. After separation, a second BH is formed leading to a BBH that will eventually coalesce through the emission of GWs.

\begin{figure}
    \centering
    \includegraphics[width=\hsize]{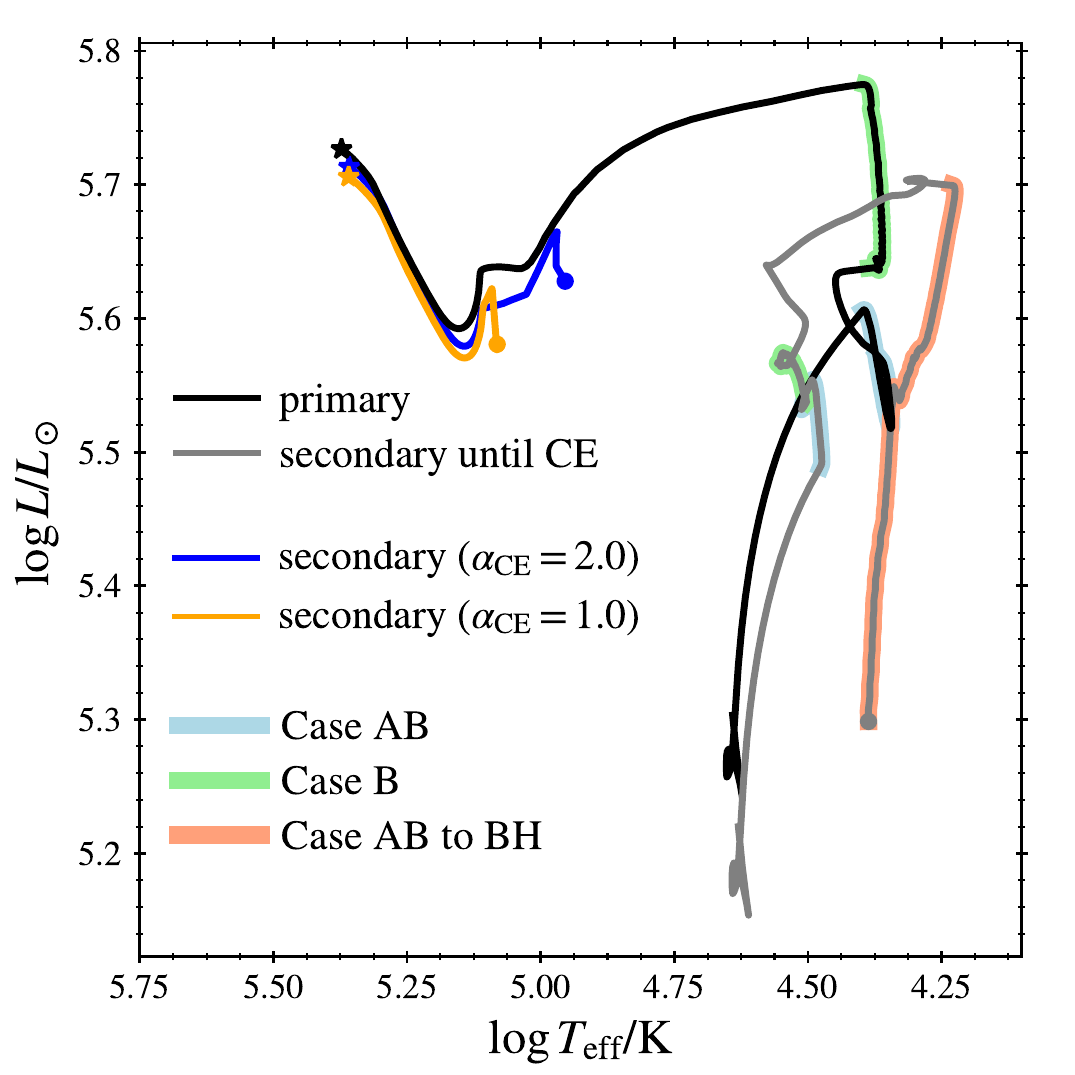}
    \caption{Full binary evolution HR diagram of the binary system considered in the example. Primary and secondary stars are born in the ZAMS (bottom right). Case AB (B) stable MT phase is indicated in light-blue (green). After them the primary moves to the left and collapses to form a first BH (black star). The secondary expands and a Case AB stable MT phase occurs (salmon colour) until an unstable CE phase is triggered (grey circle). The detach of the secondary occurs at the blue (orange) circle for $\alpha_{\rm CE}=2.0$ ($\alpha_{\rm CE}=1.0$). A second BH is formed at the top left corner (coloured stars).}
    \label{fig:hr-diagram}
\end{figure}

In Figure~\ref{fig:hr-diagram} we present the full binary evolution Hertzsprung-Russell (HR) diagram corresponding to the example binary systems. In this HR diagram, the primary and secondary stars are born in the ZAMS (in the bottom right part of the figure) and end forming BHs (in the upper left part of the plot). After the primary star expands, two stable MT phases develop: the so-called cases AB and B indicated in light-blue and green, respectively. The luminosity of the primary increases until separation occurs. Then, the primary contracts, moving to the left forming a Wolf-Rayet star. Later on, it collapses to a BH (black star in the figure). Meanwhile, the secondary continues its evolution. After leaving the MS, the secondary expands and a stable MT phase commences (Case AB to BH indicated with salmon colour). This continues until an unstable CE phase is triggered (at the grey circle). A fast out-of-equilibrium phase is developed until separation of the secondary occurs at the blue and orange circles, for $\alpha_{\rm CE}=2.0$ and 1 respectively. The secondary evolution then continues until a second BH is formed.

\begin{figure}
    \centering
    \includegraphics[width=\hsize]{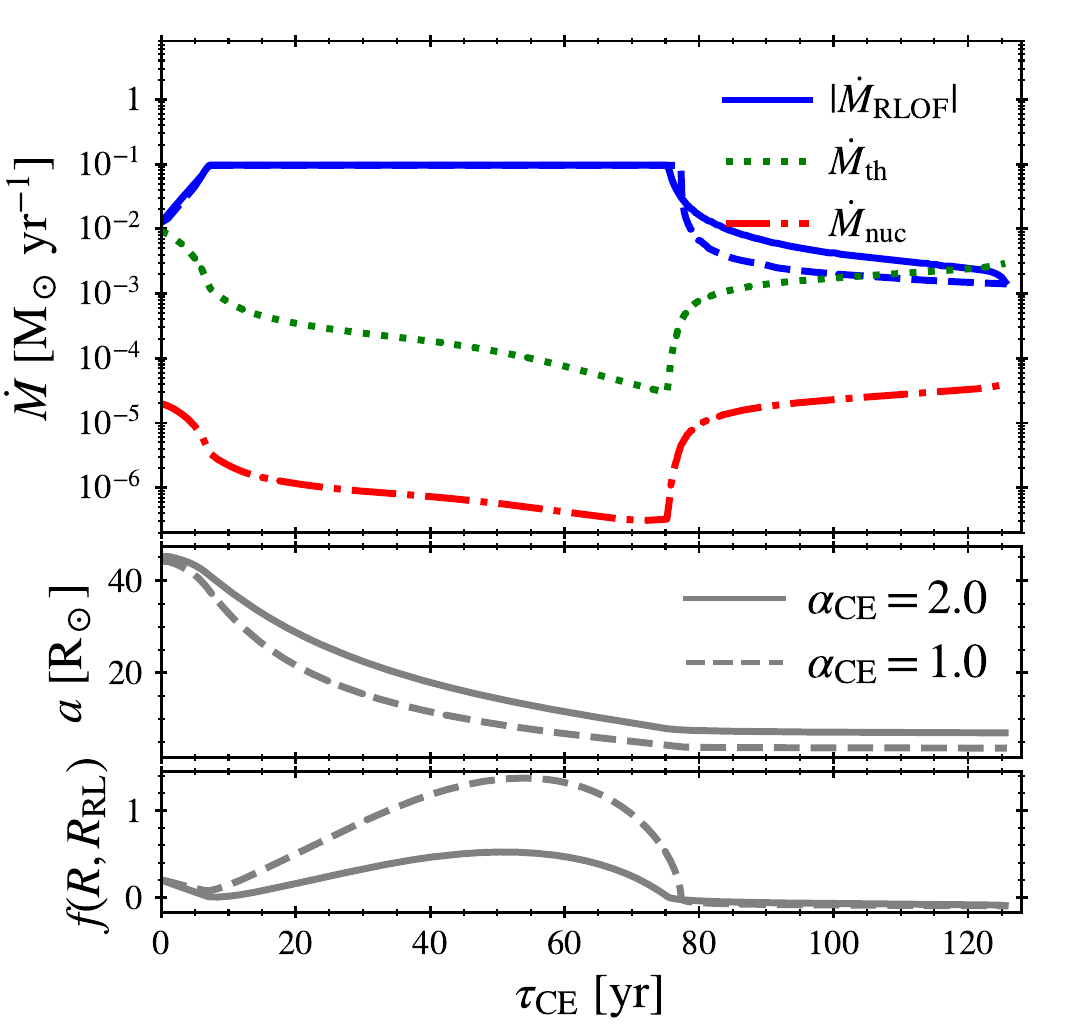}
    \caption{Evolution of binary parameters across the CE phase. On the top panel we present the evolution of the donor mass loss ($\dot{M}_{\rm RLOF}$, blue) and mass transfer scales: thermal ($\dot{M}_{\rm th}$, dotted green) and nuclear ($\dot{M}_{\rm nuc}$, dot-dashed red). On the mid panel, the separation ($a$) and on the lower panel the relative overflow ($f(R,R_{\rm RL}$)). Solid (dashed) lines represent $\alpha_{\rm CE}=2.0$ ($\alpha_{\rm CE}=1.0$).}
    \label{fig:binary-params}
\end{figure}

\begin{figure}
    \centering
    \includegraphics[width=\hsize]{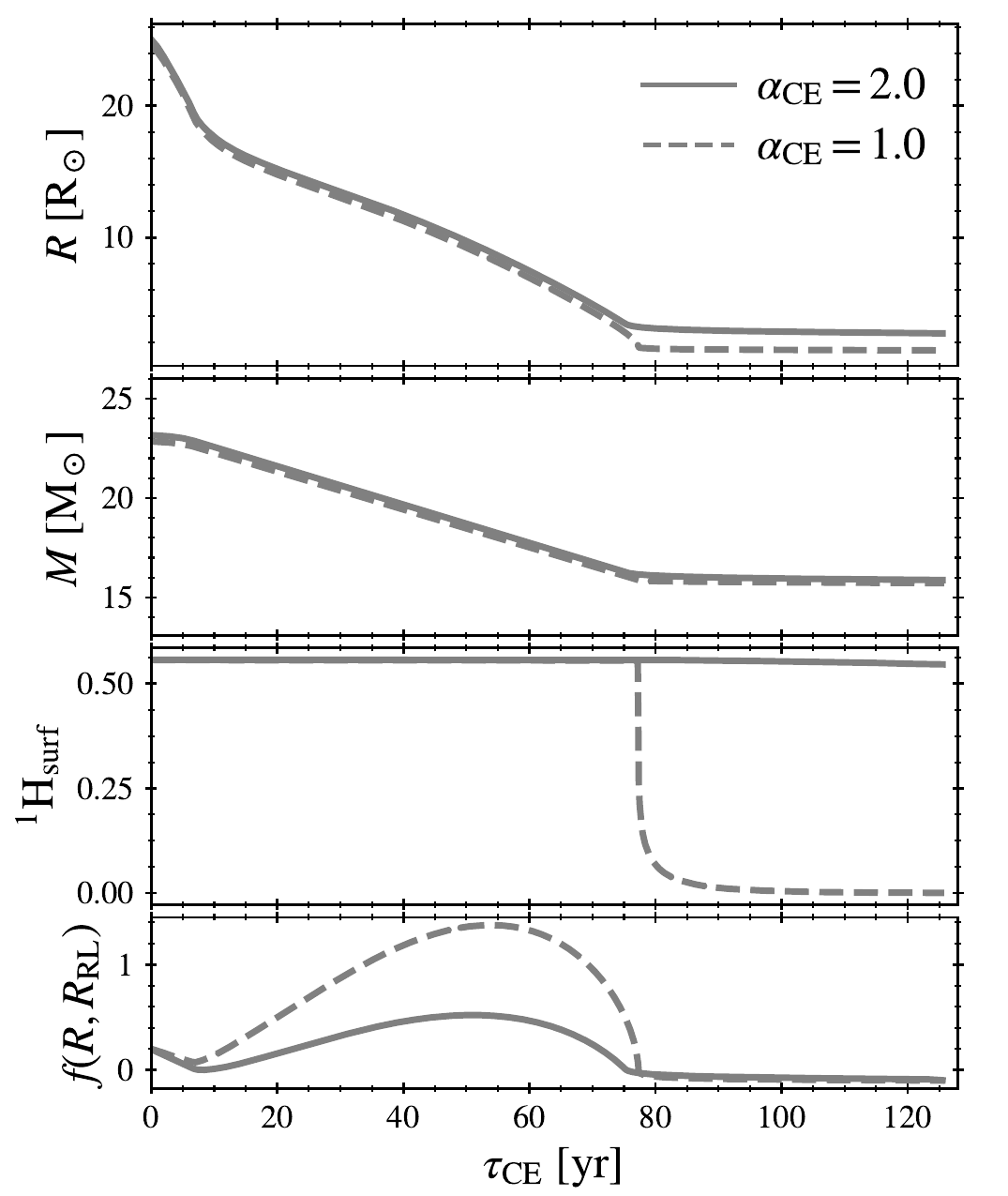}
    \caption{Evolution of donor star parameters during the CE phase. From top to bottom we present the evolution of the donor radius ($R$), donor total mass ($M$), superficial H mass fraction, and the relative overflow ($f(R,R_{\rm RL}$)). Solid (dashed) lines represent $\alpha_{\rm CE}=2.0$ ($\alpha_{\rm CE}=1.0$).}
    \label{fig:donor-params}
\end{figure}

During the short (of the order of $\sim$100~yr) unstable CE phase, the binary system evolves quickly as the orbital energy is extracted to unbind the envelope of the donor. In Figure~\ref{fig:binary-params} we show the evolution of the binary parameters during the CE phase for both CE efficiencies. In the top panel we show the evolution of the donor mass loss ($\dot{M}_{\rm RLOF}$), in the mid panel the binary separation ($a$) and in the bottom panel the relative overflow defined as $f(R,R_{\rm RL})= (R-R_{\rm RL})/R_{\rm RL}$. Furthermore, in Figure~\ref{fig:donor-params} we focus on the evolution of the donor parameters. From top to bottom we present the donor radius, total mass, superficial H mass fraction and relative overflow. 

Once the CE is triggered ($\tau_{\rm CE}=0$), the donor mass loss grows linearly for 10~yr from stable MT value to the fixed rate of $10^{-1}$~M$_\odot$~yr$^{-1}$. During the early phase, the binary separation shrinks faster than the donor radius and thus the relative overflow increases, until this effect is reversed at 50--60~yr. After that, the donor star shrinks faster until separation is reached at $\sim$80~yr, after the beginning of the CE phase. At this point the mass loss rate decreases until the thermal scale is recovered ($\dot{M}_{\rm RLOF} \approx \dot{M}_{\rm th}$) and the CE phase is finished. As a result of the CE phase, the envelope of the donor star is removed. In particular, in the $\alpha_{\rm CE}=1.0$ case, no H is left, while in the $\alpha_{\rm CE}=2.0$ case a small fraction of H remains, but its total mass decreases by $\sim$7~M$_\odot$. In both cases, a strong decrease of a factor of $\sim$10 in the orbital separation is seen, leading to an ultra-compact binary which eventually will become a BBH that will merge within the Hubble time.

\section{Merger time delay calculation}
\label{app:time_delay}

In a binary consisting of two BHs, orbital shrinking is driven by the emission of gravitation waves, which ends up with a merger of the BHs. In order to estimate the time needed for a BBH to merge after its formation, which is also known as merger time delay, we use the prescription given by \citet{1964PhRv..136.1224P}:

\begin{equation}
    \begin{split}
        t_{\rm merger} = & \; \dfrac{15}{304} \dfrac{a_{0}^{4} c^{5}}{G^{3} m_{1} m_{2} (m_{1} + m_{2})} \\
                         & \times \left[\left(1+e_{0}^{2}\right) e_{0}^{-12/19} \left(1 + \dfrac{121}{304} e_{0}^{2} \right)^{-870/2299}\right]^{4} \\
                         & \times \int_{0}^{e_{0}} {\rm d}e \dfrac{e^{29/19} \left[1+\left(121/304\right)e^{2}\right]^{1181/2299}}{\left(1-e^{2}\right)^{3/2}}
    \end{split}
\end{equation}
where $a_0$ and $e_0$ are the semi-major axis and eccentricity at BBH formation, while $m_1$ and $m_2$ are the BH masses.

\begin{figure}
    \centering
    \includegraphics[width=\hsize]{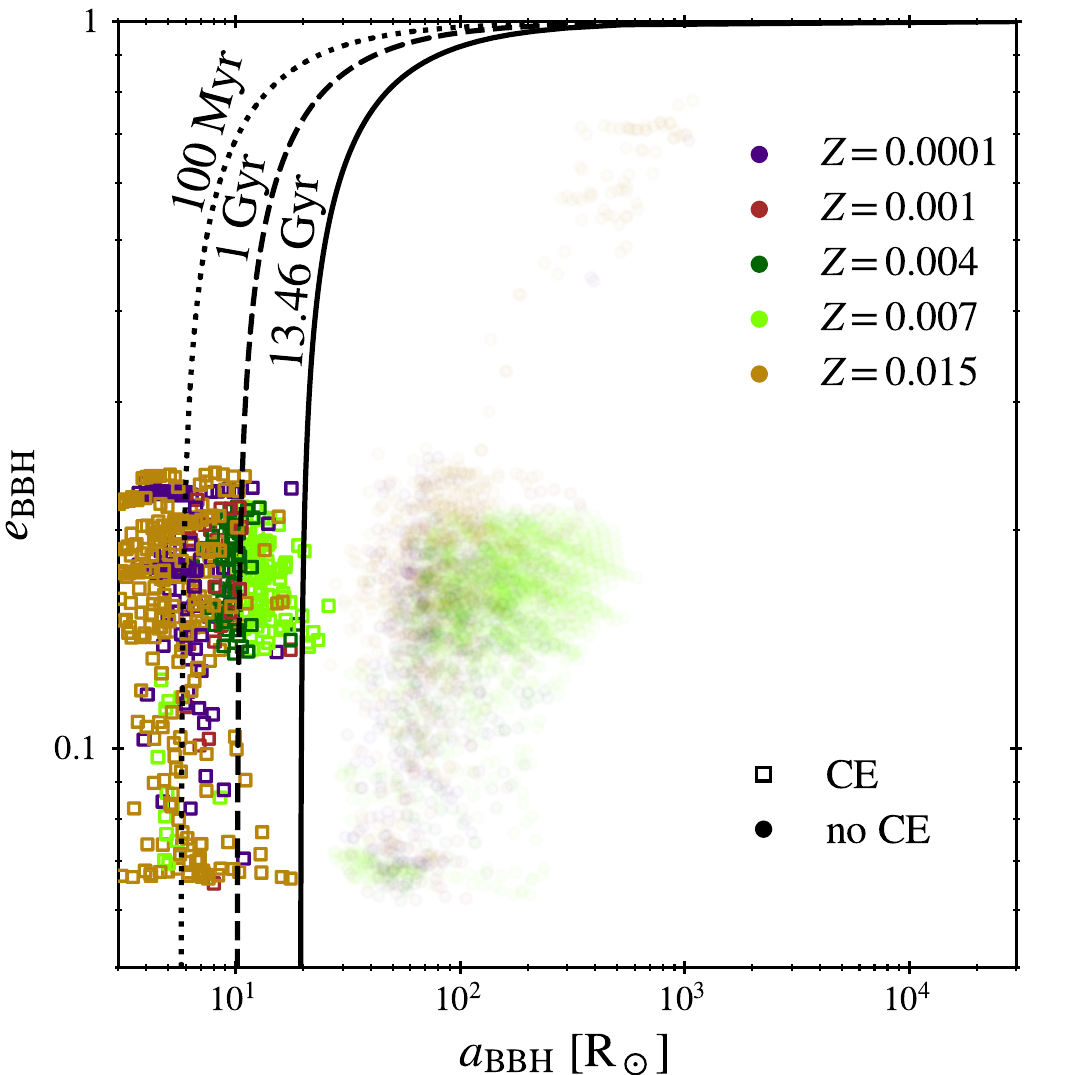}
    \caption{Final binary parameters for all our BBHs with $\mathcal{M}_{\rm chirp}$ consistent with GW151226 or GW170608. Colours indicate different metallicities (see legend). Dotted, dashed and solid black lines correspond to values of constant $t_{\rm merger}$: 100 Myr, 1 Gyr, $\tau_{\rm Hubble}$, respectively, assuming BH masses of 12.3 and 7.65 M$_\odot$.}
    \label{fig:ecc-vs-a}
\end{figure}

We show in Figure~\ref{fig:ecc-vs-a} all the BBHs found in our simulations that lie inside the 100\% C.I. of GW151226 or GW170608. Two different sub-populations can be seen: binaries that went through a CE phase and those which did not. The former have $a_{\rm BBH} < 30-40$~R$_\odot$ and most of them have a merger time delay lower than the Hubble time, while the latter have $a_{\rm BBH} \gtrsim 30-40$~R$_\odot$ and hence, merger time delays longer than the Hubble time. Therefore, the CE phase plays a key role in the formation of ultra-compact binaries which are progenitors of GW151226 and GW170608 in this evolutionary channel. Since we do not consider asymmetric BH kicks and the ejected masses in the BH prescription adopted are small (due to fallback), the BBH eccentricities arising from our simulations are generally constrained to $e_{\rm BBH} \la 0.25$. As can be seen in Figure~\ref{fig:ecc-vs-a}, for these eccentricity values, $a_{\rm BBH} \la 20$~R$_\odot$ are needed to produce BBHs with merger time delays below the Hubble time. This is because merger time delays strongly depend on the separation at BBH formation: increasing it by a factor of 10 leads to an increase in $t_{\rm merger}$ of a factor $10^4$.

\section{Dependence of the merger rates on the star formation history}
\label{app:SFRs}

\citet{2019MNRAS.490.3740N} show that the uncertainties in the metallicity evolution and star formation history can change the rates of BBH mergers. Thus, we perform the evaluation of detectable rates during O1/O2 observing runs for the progenitor population of the GW151226 and GW170608 for different SFRs and metallicity distributions. In addition to the already mentioned SFR from \citet{2004ApJ...613..200S}, we use the SFR from \citet{2017ApJ...840...39M}. For the evolution of metallicity over cosmic time, we compare \citet{2006ApJ...638L..63L} distribution with the fiducial model of \citet{2019MNRAS.490.3740N}.

\begin{figure}
    \centering
    \includegraphics[width=\hsize]{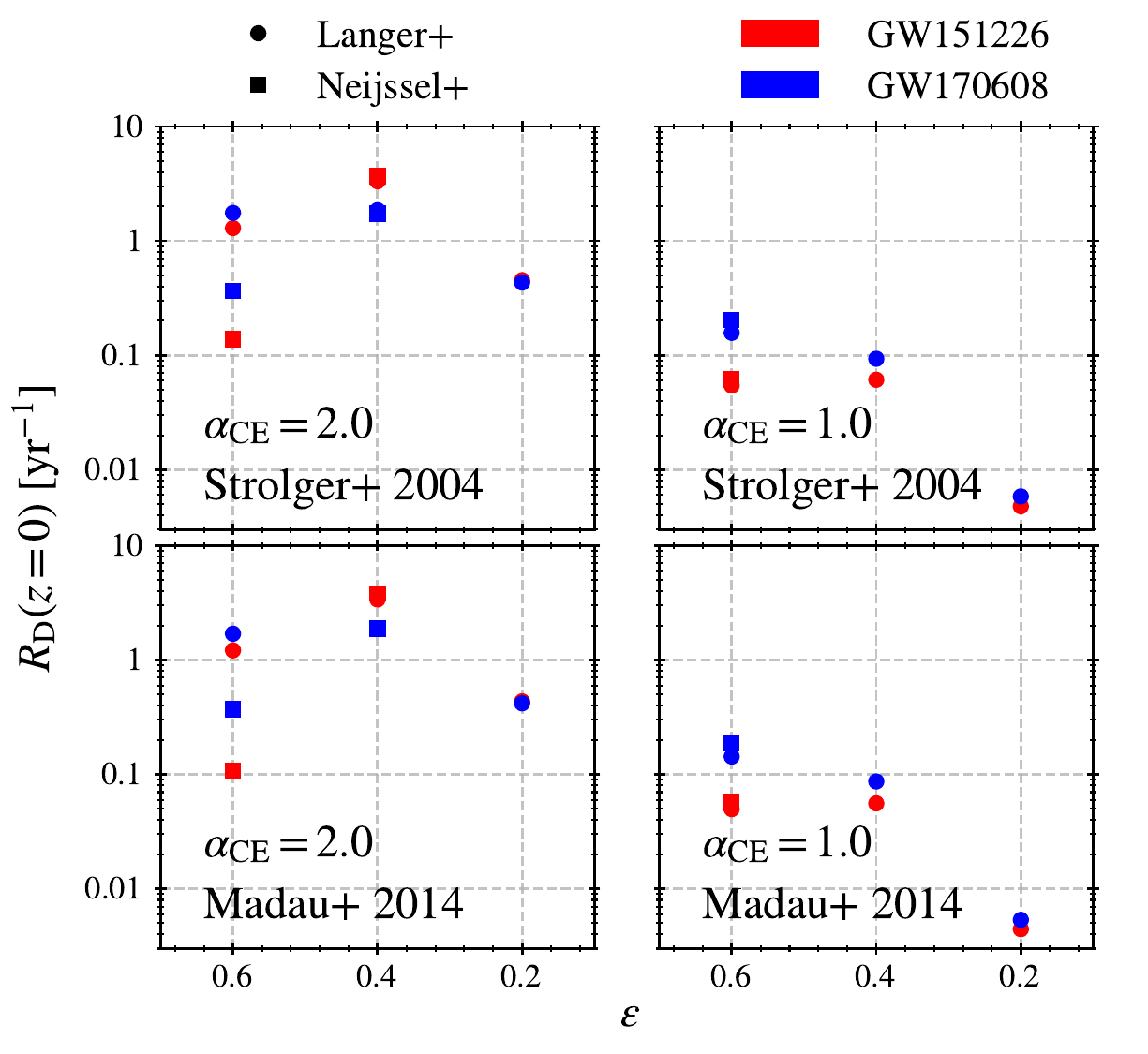}
    \caption{Total detection rates for O1 and O2 runs, $R_{D}(z=0)$, marginalised over metallicity, as a function of MT efficiency $\epsilon$ for $\alpha_{\rm CE}=2.0$ (left panel) and $\alpha_{\rm CE}=1.0$ (right panel) of events compatible with GW151226 (red) and GW170608 (blue) within their 100\% credible intervals. Circles represent detection rates assuming the metallicity evolution given in \citet{2006ApJ...638L..63L} while rectangles are the rates found using the metallicity distribution of \citet{2019MNRAS.490.3740N}.}
    \label{fig:det-rates-different-msfr}
\end{figure}

In Figure \ref{fig:det-rates-different-msfr} we present merger rates for different combinations of star-formation evolution over cosmic time. Our results show that the strongest changes in event rates are introduced by the metallicity distribution, while the different SFRs assumed produce less variations in the outcome rates. These results are similar to the ones found by \citet{2019MNRAS.488.5300C} and \citet{2019MNRAS.490.3740N}. For some cases, we find differences in the rates of more than a factor of two. In all cases, the maximum value for the detection rate remains at a level of a few per year.

\section{Black hole kicks}
\label{app:kicks}

One important and rather uncertain aspect of massive binary evolution is connected to the momentum imparted during the formation of a BH, i.e. the {\em natal kick}, similar to those that NSs receive during their formation \citep{2012ARNPS..62..407J}. This kick onto a BH could happen if, instead of having a direct collapse, a proto-NS is formed and a weak explosion leads to a large amount of mass falling back whereas a little envelope is being unbound \citep{1995MNRAS.277L..35B, 2001ApJ...554..548F}. Although the magnitude of the kick for NSs is rather well constrained from pulsar observations \citep{2005MNRAS.360..974H}, the strength of natal kicks imparted onto BHs is an open issue, as there is a debate with some arguing in favour of weak kicks \citep{2003Sci...300.1119M, 2016MNRAS.456..578M}, while others favour the opposite \citep{2012MNRAS.425.2799R,2013MNRAS.434.1355J}.

\begin{figure}
    \centering
    \includegraphics[width=\hsize]{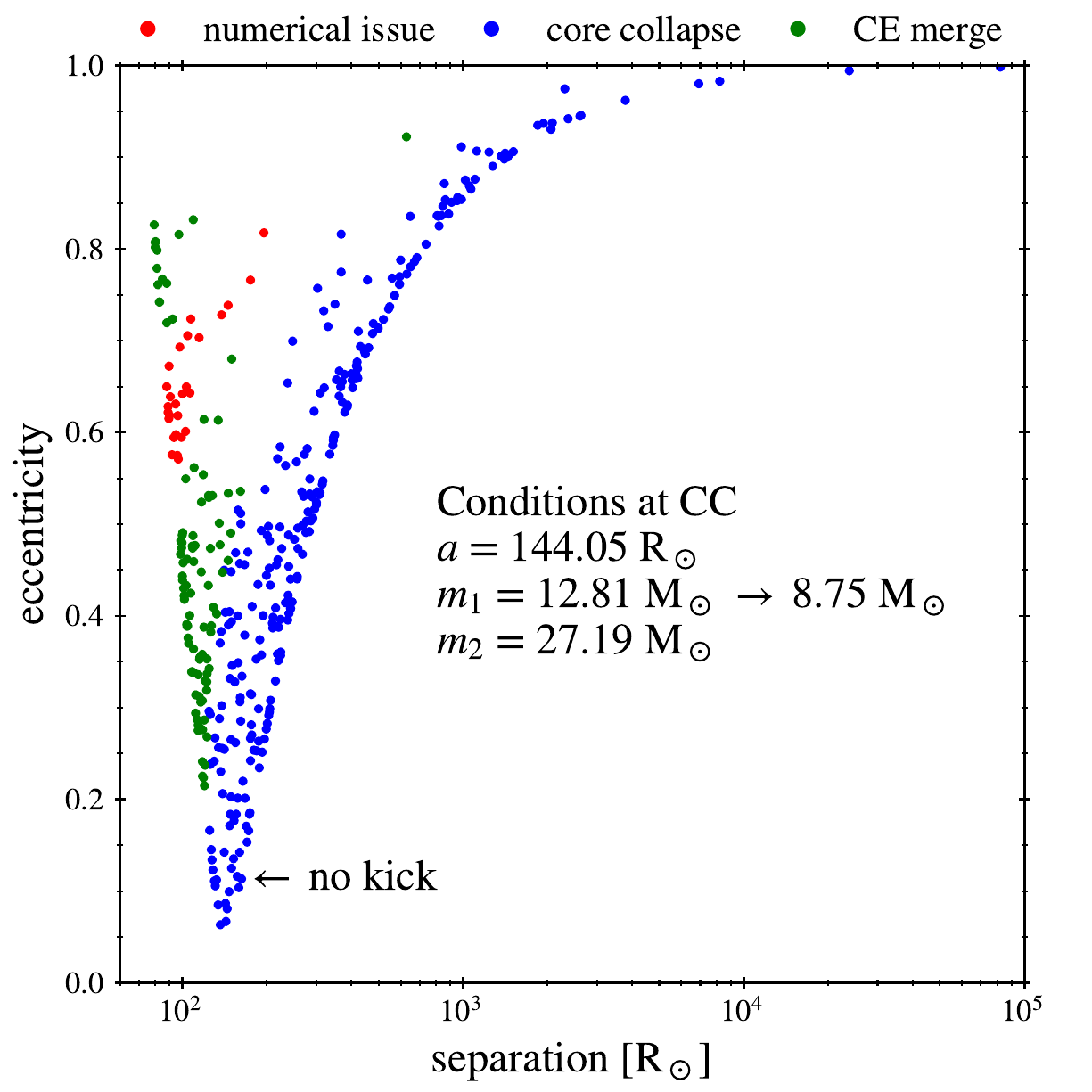}
    \caption{Binary configurations after applying a natal kick during the formation of the first BH. Masses and separations shown correspond to pre-collapse orbital parameters. Each point represents a single detailed binary evolution of the BH and its companion star. Colours show different binary outcomes: in green we represent merging binaries during a CE phase, in red we show binaries which unexpectedly end due to numerical problems, while blue points are binaries reaching the second core-collapse stage. The arrow represents the location on this plane of the binary that receives no natal kick.}
    \label{fig:binaries-after-1-kick}
\end{figure}

\begin{figure}
    \centering
    \includegraphics[width=\hsize]{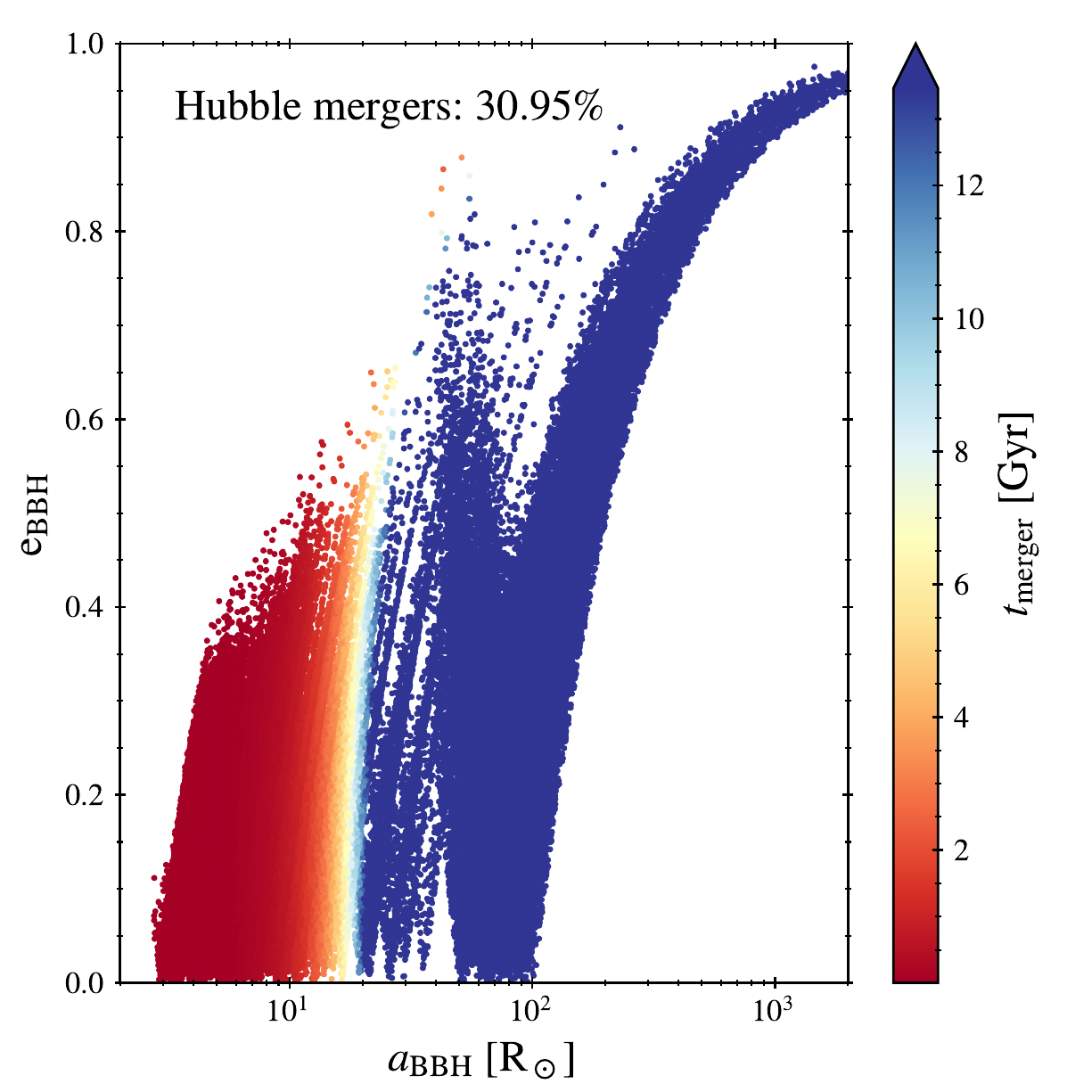}
    \caption{Binary configurations after the formation of the second BH. Each point corresponds to one of the 500 binaries randomly drawn from the blue points in Figure~\ref{fig:binaries-after-1-kick}. Colours indicate the merger delay times of the post-collapse BBHs as a result of gravitational wave radiation. After this kick, there is a~30\% chance that the BHs merge within the Hubble time.}
    \label{fig:binaries-after-2-kick}
\end{figure}

Here we present the effect introduced by considering natal kicks during the formation of a BH for each of the two core-collapse stages needed to produce a BBH system. Modelling kicks for all the first formed BHs in our sample is a difficult task in this case, as it involves drawing different velocities and directions, and then running detailed simulations for each of them. Instead, to quantify the effect of kicks, we choose one of all the simulations performed, which we identified as having consistent masses with one of the GW events under study. After the primary collapses into a BH, we randomly draw $500$ kicks from a Maxwellian distribution with a speed of 265~km~s$^{-1}$ reduced by a factor $(1 - f_{\rm fb})$, where $f_{\rm fb}$ is the fraction of mass that falls back onto the proto-NS \citep{2012ApJ...749...91F}, isotropically orientated. The post-kick binary parameters (separation and eccentricity) are updated following \citet{1996ApJ...471..352K}, assuming no interaction between the ejected mass at BH formation and the companion star. Once this initial conditions are set, each binary is evolved with {\tt MESA} as described in Sec.~\ref{sec:binary-evolution}.

In Figure~\ref{fig:binaries-after-1-kick} we show all possible post-kick binaries which remain bound after the first core-collapse (which represent $\sim$53\% of all simulated binaries). Each point in the Figure represents a binary evolved using {\tt MESA}, consisting of a BH and its companion star. We find that binaries with post-kick separations that are shorter than the one they had previous to the core-collapse tend to go through a CE phase which leads to the merger of the components, while binaries with larger separations, successfully eject the envelope, subsequently detaching and reaching a second core-collapse stage.

For each of the binaries reaching the second BH formation, we randomly apply $500$ more kicks from the same distributions mentioned before and compute the fraction of BBHs that end up merging within the Hubble time. The outcoming distribution of binary parameters at BBH formation are shown in Figure~\ref{fig:binaries-after-2-kick}. As described before, the no-kick case is a progenitor candidate to the GW events under study, as it produces a BBH system at the end of its evolution, having a merger time of~$0.2$~Gyr. For this chosen system, we obtain a $\sim$30\% probability that it would merge in less than a Hubble time if the described asymmetric kicks were applied (the remaining $\sim$70\% of the simulations are either unbound of have a much merger time longer than the Hubble time); the chances of the system being disrupted during the second core-collapse is less than $2$\%. Thus, we can expect that the addition of considering natal kicks imparted onto the BHs during both core-collapse stages would decrease the derived intrinsic rates by a factor of $\sim$3. We note however that we can not discard, as a contribution to the merger rate, the case of BBHs formed from wider stellar binaries, experiencing a fine-tuned kick, leading them to an ultra-compact remnant, and then allowing them to merge within a Hubble time.

\end{appendix}

\end{document}